\documentclass[11pt,a4paper]{article}
\usepackage{lmodern} 
\usepackage{microtype} 
\usepackage[english]{babel}

\usepackage{color,graphicx}

\usepackage{amsmath,dsfont,url,amssymb,cite}
\usepackage{slashed,cancel}
\usepackage[usenames,dvipsnames]{xcolor}
\usepackage[colorlinks=true, linkcolor=black, citecolor=ForestGreen]{hyperref}
\usepackage{mdframed}

\usepackage{subfig,authblk}
\usepackage{physics}

\usepackage{tikz}
\usetikzlibrary{positioning, arrows.meta, shapes.geometric}

\definecolor{inkblue}{RGB}{0,112,196} 	
\colorlet{blue}{inkblue}
    
\usepackage{comment}

\newcommand{\novect}[1]{#1}

\newcommand{\revchange}[1]{{#1}}

\numberwithin{equation}{section}

\begin{document}

\title{\bf Quantizing Bosonized Fermi Surfaces}

\author{Sihan Chen}
\author{Luca V.~Delacr\'etaz}
\affil{{\small \em Leinweber Institute for Theoretical Physics,\\ 
\small \em University of Chicago, Chicago, IL 60637, USA}}

\date{}
\maketitle

\begin{abstract}

Bosonization describes Fermi surface dynamics in terms of a collective field that lives on a part of phase space. While sensible semiclassically, the challenge of treating such a field quantum mechanically has prevented bosonization from providing as powerful a non-perturbative tool as in one dimension. We show that general Fermi surfaces can be exactly described by a particular $N\to \infty$ limit of a $U(N)_1$ WZW model, with a tower of irrelevant corrections. This matrix-valued description encodes the noncommutative nature of phase space, and its (solvable) strongly coupled dynamics resolves the naive overcounting of degrees of freedom of the collective field without the need to cut the Fermi surface into patches. This approach furthermore provides a quantitative tool to study power-law corrections to Fermi surface dynamics.

\end{abstract}

\maketitle

\pagebreak
\tableofcontents
\pagebreak

\section{Introduction and Summary}

Extended Fermi surfaces have a number of fascinating properties, including a continuum of gapless excitations, a landscape of possible collective modes, universal super-area law entanglement, and the existence of relevant deformations that can produce non-Fermi liquid quantum critical metals.
The extreme gaplessness of these compressible quantum phases makes them
particularly challenging to study with the conventional tools of quantum many-body physics.

Bosonization of Fermi surfaces \cite{PhysRevB.19.320,Haldane:1994,PhysRevB.48.7790,PhysRevB.49.10877,PhysRevB.51.4084,PhysRevB.52.4833,Houghton:2000bn} offers an avenue to capture some of their behavior nonperturbatively. In one spatial dimension, bosonization elegantly solves certain interacting fermion problems. In higher dimensions, it captures Landau parameters at the linearized level, simplifying the leading low-energy treatment of Fermi and non-Fermi liquids \cite{PhysRevB.48.7790,PhysRevB.49.10877,PhysRevLett.73.284, PhysRevB.73.085101}, as well as the study of collective excitations \cite{PhysRevB.48.7790,Khoo:2018nel}. Higher-dimensional bosonization further makes manifest approximate cancellations in fermion loops \cite{Delacretaz:2022ocm}, and its close connection to current algebra makes it an ideal platform to study emergent symmetries, their anomalies, and their interplay with microscopic spacetime symmetries \cite{Oshikawa:2000lrt,Else:2020jln,Delacretaz:2022ocm,Lee:2022hcm,Delacretaz:2025ifh}.

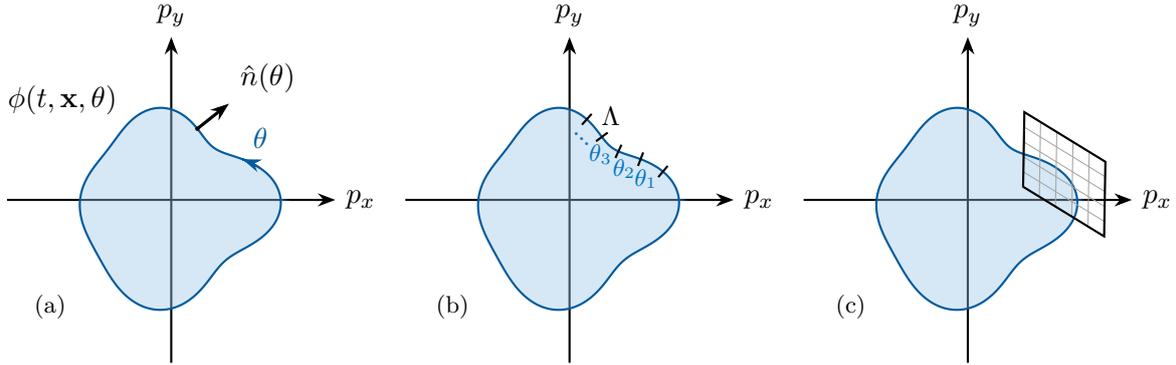
\begin{figure}
\centerline{
\captionsetup[subfloat]{captionskip=-30pt}
\subfloat[\hspace{100pt}\,]{\label{sfig_1a}
\begin{tikzpicture}[scale=1.2,>=Stealth]

  \draw[->,thick] (-1.8,0) -- (1.8,0) node[right] {$p_x$};
  \draw[->,thick] (0,-1.8) -- (0,1.8) node[above] {$p_y$};

  \filldraw[
    fill=blue!40, fill opacity=0.4,
    draw=blue!80!black, thick
  ]
    plot[smooth cycle, variable=\t, domain=0:360, samples=400]
    ({ (1 + 0.1*sin(-\t) + 0.1*cos(-3*\t) + 0.1*cos(4*\t)) * cos(\t) },
     { (1 + 0.1*sin(-\t) + 0.1*cos(-3*\t) + 0.1*cos(4*\t)) * sin(\t) });

  \pgfmathsetmacro{\thetaN}{70}
  \pgfmathsetmacro{\RN}{1 + 0.1*sin(-\thetaN) + 0.1*cos(-3*\thetaN) + 0.1*cos(4*\thetaN)}
  \pgfmathsetmacro{\pxN}{\RN*cos(\thetaN)}
  \pgfmathsetmacro{\pyN}{\RN*sin(\thetaN)}

  \pgfmathsetmacro{\nx}{0.765181*1.3}
  \pgfmathsetmacro{\ny}{0.611993*1.3}

  \draw[black, very thick, -{Stealth[length=7pt,width=5pt]}]
    (\pxN,\pyN) -- ++(0.35*\nx,0.35*\ny)
    node[above right] {$\hat n(\theta)$};

  \filldraw[black] (\pxN,\pyN) circle (0.5pt);

  \pgfmathsetmacro{\thetaStart}{20}
  \pgfmathsetmacro{\thetaEnd}{30}

  \draw[blue!80!black, thick, -{Stealth[length=7pt,width=5pt]}]
    plot[domain=\thetaStart:\thetaEnd, samples=100]
    ({ (1 + 0.1*sin(-\x) + 0.1*cos(-3*\x) + 0.1*cos(4*\x)) * cos(\x) },
     { (1 + 0.1*sin(-\x) + 0.1*cos(-3*\x) + 0.1*cos(4*\x)) * sin(\x) })
    node[above right, blue!80!black] {$\theta$};

  \pgfmathsetmacro{\thetaPhi}{120}
  \pgfmathsetmacro{\RP}{1 + 0.1*sin(-\thetaPhi) + 0.1*cos(-3*\thetaPhi) + 0.1*cos(4*\thetaPhi)}
  \pgfmathsetmacro{\pxP}{\RP*cos(\thetaPhi)}
  \pgfmathsetmacro{\pyP}{\RP*sin(\thetaPhi)}

  \node[black, above left] at (\pxP,\pyP) {$\phi(t,\vb x,\theta)$};

\end{tikzpicture}
}
\subfloat[\hspace{100pt}\,]{\label{sfig_1b}
	\begin{tikzpicture}[scale=1.2,>=Stealth]

	  \draw[->,thick] (-1.8,0) -- (1.8,0) node[right] {$p_x$};
	  \draw[->,thick] (0,-1.8) -- (0,1.8) node[above] {$p_y$};

	  \filldraw[
		fill=blue!40, fill opacity=0.4,
		draw=blue!80!black, thick
	  ]
		plot[smooth cycle, variable=\t, domain=0:360, samples=400]
		({ (1 + 0.1*sin(-\t) + 0.1*cos(-3*\t) + 0.1*cos(4*\t)) * cos(\t) },
		 { (1 + 0.1*sin(-\t) + 0.1*cos(-3*\t) + 0.1*cos(4*\t)) * sin(\t) });

	  \foreach \i/\THeta in {1/17, 2/30, 3/45, 4/63, 5/78} {
		\pgfmathsetmacro{\R}{1 + 0.1*sin(-\THeta) + 0.1*cos(-3*\THeta) + 0.1*cos(4*\THeta)}
		\pgfmathsetmacro{\px}{\R*cos(\THeta)}
		\pgfmathsetmacro{\py}{\R*sin(\THeta)}
		\pgfmathsetmacro{\dR}{-0.1*cos(-\THeta) + 0.3*sin(-3*\THeta) -0.4*sin(4*\THeta)}
		\pgfmathsetmacro{\tx}{\dR*cos(\THeta) - \R*sin(\THeta)}
		\pgfmathsetmacro{\ty}{\dR*sin(\THeta) + \R*cos(\THeta)}
		\pgfmathsetmacro{\nx}{-\ty}
		\pgfmathsetmacro{\ny}{\tx}
		\pgfmathsetmacro{\len}{sqrt(\nx*\nx + \ny*\ny)}
		\pgfmathsetmacro{\nx}{\nx/\len}
		\pgfmathsetmacro{\ny}{\ny/\len}
		\draw[black, thick] 
		  (\px-0.08*\nx,\py-0.08*\ny) -- (\px+0.08*\nx,\py+0.08*\ny);

		\ifnum\i<4
		  \node[blue] at (\px+0.2*\nx+0.1*\tx-0.04*\i+0.08,\py+0.2*\ny+0.1*\ty+0.015*\i*\i - 0.02) {\footnotesize $\theta_{\i}$};
		\fi
		\ifnum\i=4
			\node[blue] at (\px+0.2*\nx+0.1*\tx-0.01,\py+0.2*\ny+0.1*\ty+0.03) {\rotatebox{-45}{\small $...$}};
			\node[black] at (\px-0.2*\nx+0.1*\tx-0.01,\py-0.2*\ny+0.1*\ty+0.05) {\small $\Lambda$};
		\fi
	  }
	\end{tikzpicture}
}\subfloat[\hspace{100pt}\,]{\label{sfig_1c}
	\begin{tikzpicture}[scale=1.2,>=Stealth]

  \draw[->,thick] (-1.8,0) -- (1.8,0) node[right] {$p_x$};
  \draw[->,thick] (0,-1.8) -- (0,1.8) node[above] {$p_y$};

  \filldraw[
    fill=blue!40, fill opacity=0.4,
    draw=blue!80!black, thick
  ]
    plot[smooth cycle, variable=\t, domain=0:360, samples=400]
    ({ (1 + 0.1*sin(-\t) + 0.1*cos(-3*\t) + 0.1*cos(4*\t)) * cos(\t) },
     { (1 + 0.1*sin(-\t) + 0.1*cos(-3*\t) + 0.1*cos(4*\t)) * sin(\t) });

  \pgfmathsetmacro{\theta}{15}
  \pgfmathsetmacro{\R}{1 + 0.1*sin(-\theta) + 0.1*cos(-3*\theta) + 0.1*cos(4*\theta)}
  \pgfmathsetmacro{\px}{\R*cos(\theta)}
  \pgfmathsetmacro{\py}{\R*sin(\theta)}

  \begin{scope}[shift={(\px,\py)}, rotate=-\theta]
    \begin{scope}[cm={{1,-0.3,-0.2,0.8,(0,0)}}]
      \foreach \x in {-0.5,-0.3,...,0.5} {
        \draw[gray!70] (\x,-0.5) -- (\x,0.5);
      }
      \foreach \y in {-0.5,-0.3,...,0.5} {
        \draw[gray!70] (-0.5,\y) -- (0.5,\y);
      }
      \draw[thick] (-0.5,-0.5) rectangle (0.5,0.5);
    \end{scope}
  \end{scope}
	\end{tikzpicture}
}
}
\caption{\label{fig_FS} (a) Fermi surface (FS) and its bosonized degree of freedom. (b) The patch prescription separates the smooth FS in discrete patches of size $\Lambda$, and cuts off the momentum of particle hole excitations along the FS $q<\Lambda$. (c) Strongly coupled dynamics caused by the noncommutative phase-space (grid) along the FS obviates patches.}
\end{figure}

However, to turn the bosonization of Fermi surfaces into a systematic effective field theory one is faced with a challenge: making sense of fields that depend on (a part of) non-commutative phase space---reasonable objects in a semiclassical approximation---as bona fide quantum fields. Considering 2+1 dimensions for concreteness, the bosonized degree of freedom $\phi(t,\vec x,\theta)$ depends on spacetime and the Fermi surface parametrized by $\theta$. Its leading order action describes a chiral boson at each point of the Fermi surface
\begin{equation}\label{eq_S_intro}
S = -p_F \int dt d^2x d\theta \, \nabla_n \phi(\dot \phi + v_F \nabla_n \phi)+\cdots\, , 
\end{equation}
where $\nabla_n \equiv \hat n(\theta)\cdot \nabla$ is a gradient in the direction normal to the Fermi surface, see Fig.~\ref{fig_FS}. The $\cdots$ include recently identified nonlinear terms \cite{Delacretaz:2022ocm} which we will return to---they will play a key role in resolving the puzzles below. One aspect of the theory \eqref{eq_S_intro} that would appear to complicate quantization is that gradients in the direction normal to $\hat n(\theta)$, and in the $\theta$ direction, are unsuppressed.
Relatedly, at first glance this formulation also appears to vastly overcount degrees of freedom, by assigning an independent mode to every particle-hole excitation along the Fermi surface. This apparent overcounting of degrees of freedom is illustrated most clearly with a flat, discretized, Fermi surface. Consider an array of wires in the $x$ direction, each consisting of a right-moving fermion. The degree of freedom now depends on discretized space and momentum in the $y$ direction $\phi_{y,\theta}(t,x)$, with $y=1,\ldots, N_{\rm wires}$ and $\theta = 1,\ldots ,N_{\rm wires}$ running on the dual momentum lattice. If these $(N_{\rm wires})^2$ fields were independent weakly coupled modes, they would lead to a free energy (or specific heat, entropy, etc.) $\propto (N_{\rm wires})^2$ instead of $\propto N_{\rm wires}$. Current approaches to higher-dimensional bosonization propose to resolve this overcounting for general Fermi surfaces with an ad-hoc prescription: cutting the Fermi surface into patches of size $\Lambda$, and constraining the momentum of each patch field $q_y\leq \Lambda$.

The correct approach for the simple case of a flat Fermi surface, effectively a one-dimensional problem, is of course straightforward. One could bosonize each wire individually, and find an appropriate description in terms of $N_{\rm wires}$ bosons with only one of the two labels, e.g., $\phi_y(t,x)$. Alternatively, to obtain a description closer to \eqref{eq_S_intro}, one could use nonabelian bosonization \cite{Witten:1983ar} to describe the wires in terms of the chiral $U(N_{\rm wires})_1$ WZW model, whose degree of freedom $g=e^{i\phi} = \mathds 1 + i\phi(t,x) + \cdots$ is a matrix with components $\phi_{y,\theta}(t,x)$. In this case, the $\phi_{y,\theta}$ are not weakly coupled, and nonperturbative dynamics reduces the central charge realized by these $(N_{\rm wires})^2$ chiral fields to $c=\frac12  N_{\rm wires}$.  This simple example provides a hint as to how to deal with the bosonization of general Fermi surfaces, and elevate the approach to a fully fledged QFT. For general Fermi surfaces, the ellipses in \eqref{eq_S_intro} include recently found nonlinear terms that have a universal structure and capture the nonlinear response of Fermi liquids \cite{Delacretaz:2022ocm}. These nonlinear terms also turn the bosonized description into an interacting theory. 
We will show that these nonlinear terms precisely correspond to those of the $U(N_{\rm wires})_1$ WZW model for the case of a flat Fermi surface, and can similarly deal with any smooth Fermi surface. We therefore find that patches are not needed; the reduction of degrees of freedom instead occurs automatically due to strongly coupled---but solvable---dynamics along the Fermi surface. Furthermore, the patch prescription can be {\em derived}, as an alternative approach, similar to the abelian bosonization of multiple wires.

The importance of nonperturbatively treating the non-commutative structure along the Fermi surface was recognized in \cite{Polychronakos:2005br}. Our approach offers a systematic way to do so using the standard tools of (commutative) QFT. To illustrate the practical usefulness of our description, we show how it can be used to capture power-law corrections to dynamical Fermi liquid observables, focusing on the dynamic structure factor (or density two-point function) $\langle \rho\rho\rangle(\omega,q)$ and the specific heat $c_V$. These power-law corrections generalize ``beyond Luttinger'' corrections---which in 1d are well understood to arise from loop corrections in the boson description \cite{pereira2007dynamical,RevModPhys.84.1253}---to higher dimensions. While these corrections are subleading in Fermi liquid states, similar corrections are expected to play an important role for non-Fermi liquids \cite{PhysRevB.73.085101,PhysRevLett.97.226403,Delacretaz:2022ocm,Umang,Han:2023veh,Ravid:2024uqc}. For a general Fermi surface shape, the noncommutative and dispersive directions can be neatly disentangled by turning on a small magnetic field, following an approach inspired by Refs.~\cite{Ye:2024osp,Ye:2024pty}. While we focus on the $B\to 0$ limit here, our approach  may be useful to capture the local dynamics of Fermi liquids in a weak magnetic field. It also unifies Landau level bosonization \cite{PhysRevB.55.R7347} with the more geometric, Fermi surface centric, approach to bosonizing Fermi liquids in a small magnetic field \cite{Barci:2018lsp}.

\section{Coadjoint orbits for Fermi surface dynamics}\label{sec_coadj}

Consider spinless fermions $\psi(t,\vb x)$, on the lattice or in the continuum. We use continuum notation below, but the lattice perspective is useful and will be discussed as well. We will be interested in equal-time fermion bilinears, or their Wigner transform:
\begin{equation}\label{eq_bilinears}
f(\vb x,\vb p)
	\equiv \int d^dy \, e^{-i\vb y\cdot \vb p} \psi^{\dagger}(\vb x-\tfrac{\vb y}2) \psi(\vb x+\tfrac{\vb y}2)\, .
\end{equation}
Using fermion anticommutation relations $\{\psi^\dagger(\vb x_1),\psi(\vb x_2)\} = \delta^d(\vb x_1-\vb x_2)$, an arbitrary free fermion Hamiltonian $\hat H = \int_{\vb x_1 \vb x_2} H_{\vb x_2,\vb x_1} \psi^\dagger_{\vb x_1}\psi_{\vb x_2} \equiv \int_{\vb x\vb p} f(\vb x,\vb p) H(\vb x,\vb p)$ leads to the following Heisenberg equation of motion
\begin{equation}\label{eq_f_eom}
\begin{split}
0 
	&= \partial_t f - i [\hat H,f]\\
	&= \partial_t f(t,\vb x,\vb p) +i [H(\vb x,\vb p),f(t,\vb x,\vb p)]_{\rm MB}\, ,
\end{split}
\end{equation}
where in the second line we introduced the Moyal bracket between two functions of phase space
\begin{equation}\label{eq_Moyal}
[f(\vb x ,\vb p), g(\vb x ,\vb p)]_{\rm MB}
	= f(\vb x ,\vb p) 
	\, 2i \sin \tfrac{1}{2}
	\left(\stackrel{\leftarrow}{\nabla_x}\cdot\stackrel{\rightarrow}{\nabla_p} - \stackrel{\leftarrow}{\nabla_p}\cdot\stackrel{\rightarrow}{\nabla_x}\right)
		g(\vb x ,\vb p)\, .
\end{equation}
In the semiclassical limit $\partial_x \partial_p\ll 1$, this commutator becomes the Poisson bracket, and Eq.~\eqref{eq_f_eom} reduces to the collisionless Boltzmann kinetic equation, $0=\partial_t f + \{f,H\}_{\rm PB}$.%
	\footnote{Notice change of sign in the two lines of Eq.~\eqref{eq_f_eom}. While $f$ satisfies the Heisenberg equation in the QFT, in terms of the single-particle Hamiltonian function $H(\vb x,\vb p)$ it satisfies instead the Liouville equation, and can be viewed as a distribution function.}


It was found in Ref.~\cite{Delacretaz:2022ocm} that Eq.~\eqref{eq_f_eom} could be obtained as the equation of motion of the following action:
\begin{equation}\label{eq_coadj_action}
S = \int dt \Tr \left[f_0 U^{\dagger} \left(i\partial_t - H\right)U \right]
\end{equation}
The trace is over the Moyal algebra and is normalized as $\Tr g \equiv \int \frac{d^dx d^dp}{(2\pi)^d} g(x,p)$. All products are Moyal products $f \star g \equiv f \exp{\tfrac{i}{2}\left(\stackrel{\leftarrow}{\nabla_x}\cdot\stackrel{\rightarrow}{\nabla_p} - \stackrel{\leftarrow}{\nabla_p}\cdot\stackrel{\rightarrow}{\nabla_x}\right)}g$ (which can also be viewed as regular products of unitary matrices, see below). This action features an unexpected ingredient: a reference state $f_0$, and $U=e^{i\phi}$, with $\phi$ an element of the Moyal algebra, chosen such that $f = U f_0 U^\dagger$. While there is an ambiguity in choosing $f_0$ and $U$,  this freedom does not affect the equation of motion. In practice, it is useful to choose the reference state to be the expectation value of $f$, i.e. the distribution function corresponding to the Fermi surface at rest: $f_0(x,p) = \langle {\rm FS } |f| {\rm FS } \rangle =  \Theta(p\in \rm FS)$. With this choice, expanding the exponential $U = e^{i\phi}$ in $\phi$ using the Moyal algebra leads to a useful perturbative expansion \cite{Delacretaz:2022ocm}.

This action principle for Fermi liquids will be the starting point of our construction. Already classically, it produces the collisionless kinetic equation for a Fermi gas and, by adding Landau parameters, to interacting Fermi liquids. The advantage of an action principle over the equations of motion is that it offers a route toward quantization. At the quantum level, it seems like the fermion bilinears \eqref{eq_bilinears} constitute far too many degrees of freedom, if viewed as the fundamental fields. In this paper, we will show that there is nevertheless a natural way to quantize the kinetic theory of Fermi surfaces, which will produce an exact dual description of a free Fermi gas, in any dimension. In $d=1$, this will reduce to standard bosonization, to all orders in ``beyond Luttinger'' corrections.

The Moyal algebra satisfied by the $f(\vb x,\vb p)$ is sometimes called GMP algebra or $w_{\infty}$. If the fermions were discretized to live on  $N_{\rm latt}$ lattice sites, this algebra would be replaced by the Lie algebra $u(N_{\rm latt})$ of the group of unitary matrices. Of course, in 1d bosonization the continuum limit $N_{\rm latt}\to \infty$ is crucial to allow for a nontrivial reorganization of degrees of freedom. Nevertheless, we will see in Sec.~\ref{sec_flat} that in higher dimensions, discretizing the directions parallel to the Fermi surface is useful. A useful basis for the generators of $u(N)$ which makes this correspondence explicit is the 't Hooft basis $T_{\vb n}$, labeled by a vector of integers $\vb n = (n_1,n_2)$, $n_i =0,1,\ldots , N-1$. These generators satisfy
\begin{equation}\label{eq_UN_algebra}
[T_{\bf n},T_{\bf m}] = 2i \sin \left(\frac{\pi }{N} {\bf n}\times {\bf m}\right) T_{\bf n + m}\, , \qquad
\tr T_{\bf n} T_{\bf m} = N \delta_{\bf n,-\bf m}.
\end{equation}
See App.~\ref{app_q2_rhorho} for the explicit form of these generators. This reduces to the Moyal algebra \eqref{eq_Moyal} in the continuum limit, $\lim_{N\to \infty}u(N) = w_{\infty}$. More precisely, the Fourier transform of the distribution function $\int_{xp}f(x,p)e^{i(py-qx)}$ satisfies \eqref{eq_UN_algebra} with $q = \frac{2\pi}L n_1$ and $y = a n_2$.

\subsection{Recovering 1d bosonization beyond Luttinger liquids}\label{ssec_1d}

Before turning to extended Fermi surfaces in $d>1$ spatial dimensions, we will show that the coadjoint orbit approach is equivalent to the well established bosonization of Fermi (or Luttinger) liquids in $d=1$, including arbitrary dispersion relations beyond Luttinger liquids \cite{pereira2007dynamical,RevModPhys.84.1253}.

We focus for simplicity on a single right-moving Fermi point; multiple Fermi points can be treated similarly. The appropriate reference state is therefore%
	\footnote{Technically, because in this case $f_0$ does not vanish at infinity, the action as formulated in Eq.~\eqref{eq_coadj_action} is incorrect because one cannot integrate by parts, which spoils trace cyclicity ($f_0$ is not trace class). In practice, as long as at least one commutator acts on $f_0$, trace cyclicity is restored.\label{fn_f0}}
\begin{equation}
f_0(p_x) = \Theta(-p_x)\, .
\end{equation}
The dynamical degree of freedom is $f(x,p_x) = U f_0 U^\dagger$. Clearly, two unitaries $U$ and $Ue^{i\alpha}$ with $\alpha$ satisfying $[\alpha,f_0]_{\rm MB}=0$ will produce the same $f$. In the semiclassical limit, this stabilizer condition reads
\begin{equation}\label{eq_stabilizer}
0 = \partial_x \alpha(x,p_x) \partial_{p_x}f_0(p_x) \qquad \Rightarrow\qquad
	\partial_x\alpha(x,0) = 0\,.
\end{equation}
This gauge freedom can be used to put $U=e^{i\phi}$ in a useful form:%
	\footnote{Interestingly, this step would not be justified in any finite lattice \cite{Park:2023coa}. See App.~\ref{app_approaches_boso} for further discussion.}
\begin{equation}\label{eq_Poisson_gauge}
\phi(x,p) = \phi(x)\, .
\end{equation}
We will show that the resulting action \eqref{eq_coadj_action} is then identical to that obtained from traditional bosonization.

The action \eqref{eq_coadj_action} can be separated into a kinetic term and a potential (or Hamiltonian) term. Let us start with the kinetic term, which we will label ``KKS'' because it arises from the Kirillov-Kostant-Souriau symplectic form in the context of coadjoint orbits \cite{Alekseev:1988vx,Wiegmann:1989hn,PhysRevB.52.4833}%
	\footnote{Other appropriate names include Berry phase term, or WZW term---however we will see that it slightly differs from the usual WZW term in the context of nonabelian bosonization}
\begin{equation}
\begin{split}
S_{\rm KKS}
	&= \int dt \Tr \left(f_0 U^{\dagger}i \partial_t U\right) \\
	&= \int dt \Tr \left((f_0 + \tfrac{i}2[\phi,f_0] + \tfrac{i^2}{3!}[\phi,[\phi,f_0]] +\cdots ) (-\dot \phi)\right)  
\end{split}
\end{equation}
where in the second line we expanded the exponentials, used trace cyclicity, and $\dot \phi \equiv \partial_t \phi$. All commutators are Moyal brackets, for example
\begin{equation}\label{eq_adphi_f0}
[\phi,f_0]=\phi(x) 
	\, 2i \sin \left(\tfrac{1}{2}
	\stackrel{\leftarrow}{\partial_x}\stackrel{\rightarrow}{\partial_{p_x}} \right)
		\Theta(-p_{x})
	= -i \partial_x\phi(x) \delta(p_x) + \cdots\, , 
\end{equation}
where the ellipsis denotes terms involving higher derivatives of the delta-function, such as $\delta''(p_x)$---none of these terms contribute to the trace $\Tr(\cdots) \equiv \int\frac{dx dp_x}{2\pi} (\cdots)$ due to the integral over $p_x$. The same holds for higher brackets $[\phi,\cdots[\phi,f_0]]$, so that the kinetic term is simply
\begin{equation}
S_{\rm KKS}
	= - \int \frac{dt dx}{4\pi} \dot\phi\partial_x\phi \, .
\end{equation}
We already recognize the kinetic term of a chiral boson \cite{Floreanini:1987as} 
with chiral anomaly coefficient $k=1$, which describes a right-moving Weyl fermion.%
	\footnote{In other words, and to connect to various other terminologies used, this corresponds to the right-moving chiral factor of a Luttinger liquid $\mathcal L = \frac1{2\pi} \partial_t \Phi \partial_x \Theta - \frac{1}{4\pi} \left[\frac{1}{2K} (\partial_x\Theta)^2 + 2K (\partial_x\Phi)^2\right]$ with Luttinger parameter $K=1$, or a compact boson $S = - \frac{R^2}{8\pi}\int (\partial_\mu\Phi)^2$ at radius $R=2\sqrt{K} = 2$, with fields normalized as $\Phi\sim \Phi+2\pi$,  $\Theta\sim \Theta+2\pi$; in these conventions the self-dual point with $SU(2)$ symmetry is at $R=\sqrt{2}$ or $K=1/2$ \cite{Ginsparg:1988ui}.
	}

We now turn to the Hamiltonian term. We will consider translation invariant systems, so that the single particle Hamiltonian
\begin{equation}
H(x,p_x) = \epsilon(p_x)
\end{equation}
reduces to the dispersion relation of the fermions. The corresponding term in the action is therefore 
\begin{equation}\label{eq_SH_1dboso}
S_H
	= -\int dt \Tr \left(f_0 U^\dagger \epsilon U\right)
	= -\frac1{2\pi}\int dt dx \, e^{-i\phi(x)}\star h(p_x)\star e^{i\phi(x)} |_{p_x=0}\, , 
\end{equation}
where we defined the primitive of the dispersion relation%
	\footnote{In the language of the many-body equation of state, $\epsilon(p_x)$ corresponds to the chemical potential as a function of density $\rho =p_x/2\pi$, so that $h$ is energy density.}
$\epsilon(p_x)\equiv h'(p_x)$ and integrated by parts in $p$. The Moyal product, defined below \eqref{eq_coadj_action}, can be simplified in the expression above owing to the fact that it combines a function of $x$ with one of $p_x$:
\begin{equation}\label{eq_evaluating_Moyal}
\begin{split}
e^{-i\phi(x)}\star h(p_x)\star e^{i\phi(x)}
	&= e^{-i\phi(x)} e^{\frac{i}2 \stackrel{\leftarrow}{\partial_x}\stackrel{\rightarrow}{\partial_{p_x}}} h({p_x}) e^{-\frac{i}2 \stackrel{\leftarrow}{\partial_{p_x}}\stackrel{\rightarrow}{\partial_x}} e^{i\phi(x)}\\
	&= e^{-i\phi(x)} h({p_x}+ \tfrac{i}{2}\stackrel{\leftarrow}{\partial_x} - \tfrac{i}2\stackrel{\rightarrow}{\partial_x})e^{i\phi(x)}\\
	&= \lim_{\delta \to 0}h({p_x} + i\partial_\delta) e^{-i[\phi(x+\delta/2)-\phi(x-\delta/2)]}
\end{split}
\end{equation}
The Hamiltonian piece of the action is therefore simply
\begin{equation}\label{eq_SH_1d_neat}
S_H = - \int\frac{dt dx}{2\pi} e^{-i\phi(x)} h(-i\partial_x) e^{i\phi(x)}\, .
\end{equation}
Taylor expanding the function $h$ around the origin and collecting both kinetic and potential terms finally leads to the action
\begin{equation}\label{eq_abelian_perturbation}
S = - \int \frac{dt dx}{2\pi} \frac12 \partial_x\phi (\dot \phi + \epsilon' \partial_x\phi) 
	+ \frac{\epsilon''}{3!} (\partial_x\phi)^3 
	+ \frac{\epsilon'''}{4!} \left[(\partial_x\phi)^4 + (\partial_x^2 \phi)^2\right] + \cdots\, .
\end{equation}
The first term corresponds to the action of a right-moving chiral boson with velocity $v_F = \epsilon'=h'' \equiv \frac{d}{dp}\epsilon(p)|_{p=p_F=0}$. The next terms correspond to the leading irrelevant corrections to the Luttinger liquid due to a nonlinear fermion dispersion $\epsilon'(p_x)\neq$ const. We show in App.~\ref{app_approaches_boso} that these match to all orders with the corrections obtained in conventional 1d bosonization.

While we have considered spinless fermions, this formalism allows to introduce spin as well. In fact, spin would be treated identically to several independent wires, which are discussed in the next section.

\section{Flat Fermi Surfaces}\label{sec_flat}

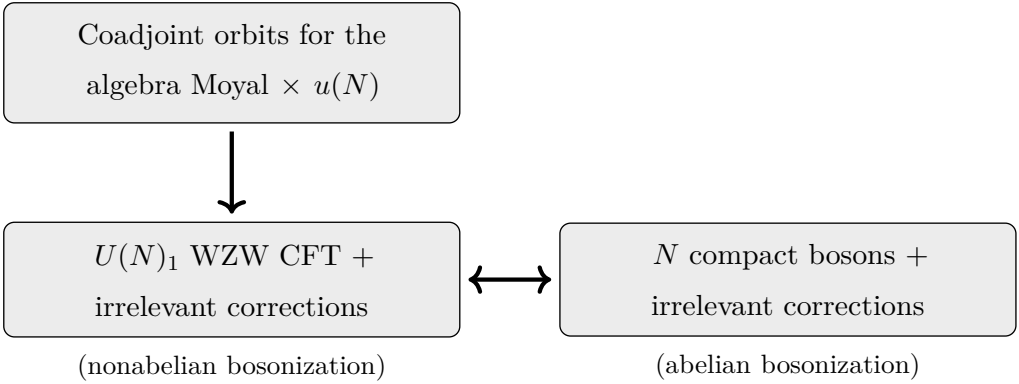
\begin{figure}
\begin{center}
    \begin{tikzpicture}[
        node distance=1.3cm,
        every node/.style={draw, line width=0.5pt, fill=lightgray!30, text width=6cm, align=center, rounded corners, inner ysep=8pt, inner xsep=0pt},
        every path/.style={ultra thick, -latex}
    ]
        \node (coadjoint) {Coadjoint orbits for the algebra $\text{Moyal} \times u(N)$};
        \node (wzw) [below=of coadjoint] {$U(N)_1$ WZW CFT + irrelevant corrections};
        \node (bosons) [right=of wzw] {$N$ compact bosons + irrelevant corrections};
        
        \node[below=of wzw, yshift=39pt, draw=none, fill=none] {\small (nonabelian bosonization)};
        \node[below=of bosons, yshift=39pt, draw=none, fill=none] {\small (abelian bosonization)};
        
        \draw[shorten >=3pt, shorten <=3pt, ->] (coadjoint) -- (wzw);
        \draw[shorten >=3pt, shorten <=3pt, <->] (wzw) -- (bosons);
    \end{tikzpicture}
\end{center}
\vspace{-20pt}
\caption{\label{fig_bosonizations} Relation between various approaches to Fermi surface bosonization. $N$ can play the role of spin or fermion flavors in 1d, number of wires for flat Fermi surfaces, or number of magnetic flux for general 2d Fermi surfaces.}
\end{figure}

As a first step towards establishing a quantum nonlinear bosonization of general Fermi surfaces, we consider flat Fermi surfaces. The absence of dispersion parallel to the Fermi surface implies that flat Fermi surfaces are effectively one dimensional systems---they are therefore simple to describe using the conventional techniques of 1d bosonization, as has been long appreciated \cite{PhysRevB.50.11446,PhysRevLett.85.2160}.%
	\footnote{See also Ref.~\cite{Lee:2022hcm} for a recent extension of these techniques beyond strictly flat chiral Fermi surfaces.}
For example, one can consider a collection of decoupled 1d wires, and take the continuum limit $N_{\rm wires}\to \infty$. The wires can either be bosonized independently (abelian bosonization), or equivalently as a whole in terms of a $U(N_{\rm wires})_1$ WZW model (nonabelian bosonization). In this section, we will see how the coadjoint orbit description reduces to nonabelian bosonization for flat Fermi surfaces. This chain of logic, which we will be able to use to tackle general Fermi surfaces as well, is illustrated in Fig.~\ref{fig_bosonizations}. Of course, in the context of flat Fermi surfaces it is straightforward to directly bosonize fermions using abelian or nonabelian bosonization. The point of this exercise is to first test and illustrate our approach in a simple context.

\subsection{From coadjoint orbits to nonabelian bosonization}\label{ssec_coadj_to_WZW_flat}

We slightly generalize the construction in Sec.~\ref{sec_coadj} by studying the algebra of bilinears made out of $N$ complex fermions
\begin{equation}\label{eq_bilinear_flat}
f_A(\novect x,\novect p)
	\equiv \int dy \, e^{-i\novect y \novect p} \psi^{\dagger}(\novect x-\tfrac{\novect y}2) T_A \psi(\novect x+\tfrac{\novect y}2)\, , 
\end{equation}
where $T_A$ runs over the elements of $u(1)\oplus su(N) = u(N)$, and can for example be taken to be the 't Hooft generators $T_{\vb n}$ in \eqref{eq_UN_algebra}. $N$ can represent the number of wires $N=N_{\rm wires}$ in a coupled wire construction, or fermion spin or flavor in a one-dimensional system. To study flat Fermi surfaces, we take the continuum limit from the start in the $x$-direction, and keep the other directions discrete for the intermediate steps. The fermion bilinear is therefore an element of the algebra%
    \footnote{While products of Lie algebras are typically not Lie algebras, for unitaries ones has $u(N)\otimes u(M) = u(NM)$}
\begin{equation}
f 
	\in  w_{\infty} \otimes u(N)\, .
\end{equation}
We follow the approach of Sec.~\ref{sec_coadj} to obtain the coadjoint orbit action. Focusing on a single right moving Fermi point, we take the reference state to be
\begin{equation}
f_0(x,p) = \Theta(-p) \otimes \mathds 1  
	\quad \in\quad  w_{\infty} \otimes u(N)\, .
\end{equation}
An element of the stabilizer $[\alpha,f_0]=0$ must satisfy
\begin{equation}
[\Theta(-p),\alpha^A(x,p)]_{\rm MB} = 0 \qquad \forall A\, .
\end{equation}
The prescription \eqref{eq_Poisson_gauge} thus now amounts to fixing
$\phi^A(x,p) \to \phi^A(x)$.
The action is again 
\begin{equation}\label{eq_coadj_flat}
S = \int dt \Tr \left[f_0 U^{-1} (i \partial_t - H)U\right]\, ,
\end{equation}
where $U=e^{i\phi}$, and $f_0,\,H \in  w_{\infty} \otimes u(N)$, and the trace is over elements of the algebra in the fundamental representation. We choose a Hamiltonian that preserves $U(N)$ and translation symmetry
\begin{equation}
H  = \epsilon(p) \otimes \mathds 1 \, ,  
\end{equation}
and consider first a relativistic dispersion relation, $\epsilon(p) = v_F p$. We will show that for such a dispersion relation  this action is {\em equal} to that of the chiral $U(N)_1$ WZW CFT \cite{Witten:1983ar,Polyakov:1984et,Sonnenschein:1988ug,Stone:1989jc}
\begin{equation}\label{eq_chiral_WZW}
\begin{split}
S 
	&= -\frac1{4\pi}\int {\rm tr} \left(\partial_x g (\partial_t + v_F \partial_x)g^{-1}\right)
	- \frac{1}{12\pi} \int {\rm tr} \left((g^{-1}dg)^3\right)\\
	&\equiv S_{\rm kin} + v_F S_H + S_{\rm WZW}\, .
\end{split}
\end{equation}
In this expression, $g(x) = e^{i\phi(x)} \in U(N)$ is the exponential of $\phi(x)$ viewed as an element of $u(N)$.%
    \footnote{\revchange{It therefore differs from the exponential $U = e^{i\phi}$ of an element of $u(N) \otimes w_\infty$. Of course, after gauge fixing \eqref{eq_Poisson_gauge}, both expressions are equal $U=g$. We nevertheless keep the notation $U$ when products are to be evaluated in the full $u(N) \otimes w_\infty$ algebra. For example, for $H = \mathds 1\otimes\epsilon(p)$ we write $U^{-1} H U = g^{-1}\star \epsilon \star g$.}}

To relate the coadjoint orbit action \eqref{eq_coadj_flat} to the $U(N)_1$ model, we can evaluate the Moyal products as in \eqref{eq_evaluating_Moyal}:
\begin{equation}\label{eq_f_nonab}
\begin{split}
f -f_0= g \star f_0 \star g^{-1} - f_0
	&= g(x) \left[f_0(p+ \tfrac{i}{2}\stackrel{\leftarrow}{\partial_x} - \tfrac{i}2\stackrel{\rightarrow}{\partial_x}) - f_0(p)\right]g^{-1}(x)\\
	&\simeq \delta(p)gi \partial_xg^{-1} - \frac12 \delta'(p) \partial_xg \partial_x g^{-1} +\cdots \, , 
\end{split}
\end{equation}
where in the second line we used $f_0(p) = \Theta(-p)$ and expanded. The $\cdots$ involve terms with more derivatives acting on the delta-function. The density operator therefore agrees with the one of the WZW model
\begin{equation}\label{eq_j_nonab}
j^0 = \int \frac{dp}{2\pi} (f-f_0) = \frac{i}{2\pi} g \partial_x g^{-1}\, .
\end{equation}
Similarly, the Hamiltonian corresponding to an arbitrary dispersion relation $\epsilon(p) = h'(p)$ is
\begin{equation}\label{eq_SH_nonab}
S_H
	= -\int dt \Tr \left[(f-f_0) \epsilon\right]
	= -\int \frac{dt dx}{2\pi} \tr \left(g h(-i \partial_x) g^{-1}\right) + \hbox{const}\, ,
\end{equation}
where tr is the $u(N)$ trace and $\Tr = \int_{xp}$tr that of the full algebra. For a linear dispersion $\epsilon(p) = v_F p$, we indeed recover the Hamiltonian part of \eqref{eq_chiral_WZW}.

We now turn to the KKS term. Let us express it in terms of $f-f_0$, which---unlike $f$ or $f_0$---has compact support, allowing for the use trace cyclicity (see Footnote \ref{fn_f0}). First, the even in $\phi\to -\phi$ part of the KKS term can be written
\begin{equation}
\begin{split}
S_{\rm KKS,\, even}
	&= \frac12 \int dt \Tr \left(f_0 \left(U^{-1}i \partial_t U + Ui \partial_t U^{-1}\right)\right)\\
	&=\frac12\int dt \Tr \left((f_0-f) Ui \partial_t U^{-1}\right)\\
	&= \int \frac{dt dx}{4\pi} \tr \left( \partial_xg \partial_t g^{-1}\right)
	= S_{\rm kin}\, ,
\end{split}
\end{equation}
where we used \eqref{eq_f_nonab} in the last line. This agrees with the kinetic term in Eq.~\eqref{eq_chiral_WZW}, which is also even under $\phi\to -\phi$ (or $g\to g^{-1}$). Next, to similarly isolate a $f-f_0$ factor in the odd part of the KKS term, we need to write it in terms of an integral in one higher dimension $\int dt \Tr f_0 U^{-1}\partial_t U = \int dt ds \Tr f_0[U^{-1}\partial_t U, U^{-1}\partial_s U]$. One then has
\begin{equation}
\begin{split}
S_{\rm KKS,\, odd}
	&=\frac{i}2 \int dtds \Tr \left(f_0 [U^{-1}\partial_t U,U^{-1}\partial_s U]\right)- (U\to U^{-1})\\
	&=\frac{-i}2\int dtds{\rm Tr} \left((f_0-f) [\partial_t UU^{-1},\partial_s UU^{-1}]\right)\\
	&= -\int \frac{dt dx ds}{4\pi} \tr \left(g^{-1}\partial_x g [g^{-1}\partial_t g,g^{-1}\partial_s g]\right)
	=\frac{-1}{12\pi}\int {\rm tr} (g^{-1}dg)^3 = S_{\rm WZW}\, .
\end{split}
\end{equation}
We therefore find that the KKS term in \eqref{eq_coadj_flat} is equal to $S_{\rm kin} + S_{\rm WZW}$ in \eqref{eq_chiral_WZW}.

\subsection{General fermion dispersion}

For a linear dispersion $\epsilon(p)=v_F p$, we have found that the coadjoint orbit action reduces to the $U(N)_1$ WZW model. For a general dispersion $\epsilon(p)$, the Hamiltonian term in the action takes the form \eqref{eq_SH_nonab}
\begin{equation}\label{eq_SH_nonab_again}
S_H
	= - \int dt \Tr \left(f_0 U^\dagger \epsilon U\right)
	= - \int \frac{dt dx}{2\pi} {\rm tr} \left(g^{-1} h(-i \partial_x)g\right)\, ,
\end{equation}
where $h'(p) = \epsilon(p)$.  We show in App.~\ref{app_approaches_boso} using the nonabelian bosonization dictionary that this indeed matches the Hamiltonian of $N$ Weyl fermions, i.e.
\begin{equation}
\frac{1}{2\pi} {\rm tr} \left( g^{-1} h(-i \partial_x) g\right)
	= \psi^\dagger_i \epsilon(-i \partial_x)\psi_i + \hbox{total derivative}\, .
\end{equation}
A nonlinear dispersion therefore leads to irrelevant corrections to the $U(N)_1$ WZW CFT. The first few can be expanded as in \eqref{eq_abelian_perturbation} and are
\begin{equation}\label{eq_SH_expand_flat}
\delta S_H = -\int \frac{dt dx}{2\pi} 
{\rm tr}\left[	\frac{\epsilon''}{3!} \left(2\pi j_x\right)^3
+	\frac{\epsilon'''}{4!} \left( (2\pi j_x)^4 + (2\pi \partial_x j_x)^2 \right)
+ 	\cdots\right]\, .
\end{equation}
Products of operators should be understood as being normal ordered. The $U(N)$ current is $j_x = \frac{1}{2\pi } g i \partial_x g^{-1}$; in the abelian case ($N=1$), $j_x = \frac1{2\pi}\partial_x \phi$ and one recovers Eq.~\eqref{eq_abelian_perturbation}.

\subsection{Importance of non-perturbative dynamics}

The nonperturbative dynamics of the WZW model resolves the naive overcounting of degrees of freedom of higher-dimensional bosonization of Fermi surfaces. Indeed, while the action \eqref{eq_coadj_flat} is a theory of $N^2$ bosons $\phi_{ij}(t,x)$, they are strongly coupled and their central charge is not $N^2$ but
\begin{equation}
c_{U(N)_k} = c_{U(1)} + c_{SU(N)_k}
	= 1 + k \frac{N^2-1}{N+k} \xrightarrow{k=1} N\, ,
\end{equation}
or $c = \frac12 N$ for the chiral model. Taking the thermodynamic limit in this wire construction, $N = L_y/a \to \infty$, the reduction of degrees of freedom $N^2\to N$ implies that free energy and specific heat are extensive as expected (and not superextensive). This reduction is also crucial to capture power-law corrections coming from irrelevant corrections \eqref{eq_SH_expand_flat}.

While our discussion so far merely revisits well-known dualities in 1+1d CFT, we will see that a similar reduction of degrees of freedom is at play for general smooth Fermi surfaces.

\subsection{Abelianization}\label{ssec_abelianization}

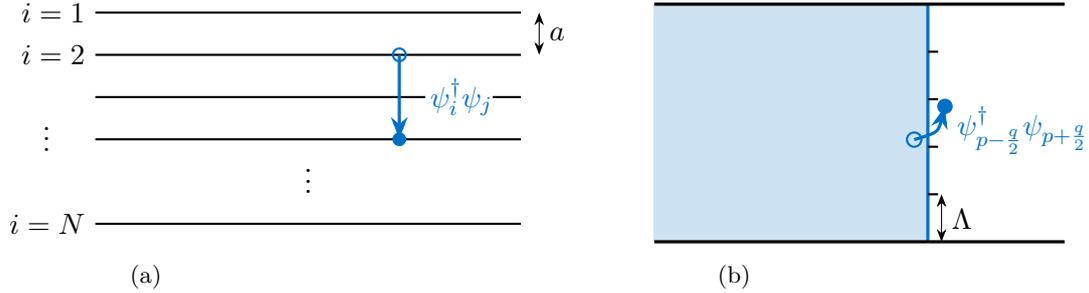
\begin{figure}
\begin{center}
\subfloat[\hspace{100pt}\,]{\label{sfig_3a}
\begin{tikzpicture}[scale=0.8,>=Stealth]

  \def\L{7} 
  \def\a{0.7} 

  \foreach \i in {1,2,3,4} {
    \draw[thick] (-\L/2, -\i*\a) -- (\L/2, -\i*\a);
    \ifnum\i<3
		\node[left] at (-\L/2, -\i*\a) {$i=\i$};
	\fi
  }

  \node at (0,-4.8*\a) {$\vdots$};
  \node at (-0.5*\L - 0.8,-3.8*\a) {$\vdots$};

  \def\N{6}
  \draw[thick] (-\L/2, -\N*\a) -- (\L/2, -\N*\a);
  \node[left] at (-\L/2, -\N*\a) {$i=N$};

  \draw[<->] (\L/2+0.3, -1*\a) -- (\L/2+0.3, -2*\a)
    node[midway,right] {$a$};

  \def\i{2}
  \def\j{4}
  \def\xpos{1.5} 

  \coordinate (pi) at (\xpos, -\i*\a);
  \coordinate (pj) at (\xpos, -\j*\a);

  \draw[very thick,->,blue] (pi) -- (pj);
  \node[blue,fill=white] at (\xpos+1.4*\a,-0.5*\a*\i-0.5*\a*\j) {\!$\psi_i^\dagger \psi_j$\!\!};

  \filldraw[blue] (pj) circle (3pt); 
  \draw[blue,thick] (pi) circle (3pt); 

\end{tikzpicture}
}\qquad
\subfloat[\hspace{100pt}\,]{\label{sfig_3b}

\begin{tikzpicture}[scale=0.85,>=Stealth]

  \def\pymax{3.5}

  \fill[blue!20] (-4,0) rectangle (0,\pymax);

  \draw[very thick,blue] (0,0) -- (0,\pymax);

  \draw[very thick,black] (-4,0) -- (2,0);
  \draw[very thick,black] (-4,\pymax) -- (2,\pymax);

  \foreach \k in {1,2,3,4} {
    \draw[thick] (0,\k*\pymax/5) -- ++(0.15,0); 
  }

  \draw[<->] (0.2,0) -- (0.2,\pymax/5)
    node[midway,right] {$\Lambda$};

  \coordinate (hole) at (-0.2, \pymax*2.15/5);
  \coordinate (particle) at (0.25, \pymax*2.85/5);

  \draw[very thick,->,blue] (hole) .. controls (0.15, \pymax*2.3/5) .. (particle)
    node[midway,right] {\ $\psi^{\dagger}_{p_y-\frac{q_y}2} \psi_{p_y+\frac{q_y}2}$};

  \draw[blue,thick] (hole) circle (3pt); 
  \filldraw[blue] (particle) circle (3pt); 

\end{tikzpicture}
}
\end{center}
\vspace{-10pt}
\caption{\label{fig_flatFS} 
(a) $N$ wires, each containing a chiral fermion. Fermion bilinears $\psi^\dagger_i\psi_j$ form the $u(N)$ algebra, which becomes the Moyal $w_\infty$ algebra in the continuum limit $N\to \infty$. (b) Corresponding flat Fermi surface. The patch prescription can be derived by abelianizing our approach, using the $u(1)^N$ subalgebra spanned by the bilinears $f(q,p) = \psi^{\dagger}_{p-\frac{q}2} \psi_{p+\frac{q}2}$ satisfying \eqref{eq_u1N_patch}.}
\end{figure}

Before turning to general Fermi surfaces, we show in the simpler context of flat Fermi surfaces how the patch (or ``pill box'') prescription that is commonly used in higher-dimensional bosonization \cite{PhysRevB.48.7790,PhysRevB.49.10877,Houghton:2000bn} can be derived from our approach. First, note that if each wire had been independently bosonized (abelian bosonization), the description would involve $N$ abelian bosons $\phi_i(t,x)$, with $i$ running over the wire label. From the perspective of nonabelian bosonization, this description arises because level-1 WZW models have a vertex representation where a free compact boson is associated to every Cartan generator, in this case spanning the $u(1)^N$ subalgebra (maximal torus) of $u(N)$ \cite{francesco2012conformal}; we will refer to this as the {\em abelianization} of the description. There are many such subalgebras, all related by $U(N)$ conjugation. However, given our phase-space interpretation of $U(N)$, local operators in $d$ spatial dimensions will have different expressions depending on the $u(1)^N$ subalgebra that is chosen. The choice that leads to the usual abelian bosonization of individual wires is $f_i = \psi^\dagger_i \psi_i$, $i=1,\ldots, N$. A different choice will lead to the patch prescription: consider
\begin{equation}\label{eq_patch_gen}
f(q_y,p_y)
	= \int dy\, e^{-iq_y y} f(y,p_y)
	= (\psi_{p_y-\frac{q_y}2})^\dagger\psi_{p_y+\frac{q_y}2}\, . 
\end{equation}
Here all momenta and coordinates refer to the direction along the Fermi surface---the dispersive direction has been dropped for clarity. Consider a momentum scale $\Lambda$, which will correspond to the patch size. The following set of operators commute:
\begin{equation}\label{eq_u1N_patch}
\revchange{
f(q_y,p_y)\, ,\qquad 0\leq q_y < \Lambda\, ,\quad  p_y = \Lambda \mathbb Z\, . }
\end{equation}
These correspond to particle-hole excitations within a single patch, and centered at the middle of the patch, see Fig.~\ref{sfig_3b}.
In the discretized set-up with $N$ wires that we were considering above, this corresponds to $N$ operators spanning a maximal torus $u(1)^N$: $p\in\Lambda \{0,1,\ldots N_{\rm patches}-1\}$ with $\Lambda N_{\rm patches} = \frac{2\pi}{a}$, and $q = \frac{2\pi}{a} \{0,1,\ldots \frac{N}{N_{\rm patches}} - 1\}$, where $a$ is the separation between the wires.

Let us now determine the action in this abelianized description. Given a choice of a $u(1)^N$ subalgebra with generators $T_i$, $i=1,2,\ldots,N$, it is obtained by taking the group element of the WZW model $g \to e^{i\phi_i(t,x)T_i}$.%
	\footnote{More precisely, one can expand the currents in the Weyl-Cartan basis and show that only the Cartan currents contribute to the stress tensor \cite{francesco2012conformal}.}
The generators $T_{(q,p)}$ corresponding to \eqref{eq_patch_gen} have matrix elements $(T_{(q,p)})_{y_1y_2} = e^{-iq(y_1+y_2)/2} e^{-i(y_1-y_2)p}/N$, with traces given by
\begin{equation}
\tr T_{(q_1,p_1)}\cdots T_{(q_n,p_n)}
	= \delta_{\Sigma_i q_i,0} \delta_{p_1,p_2}\delta_{p_1,p_3}\cdots \delta_{p_1,p_n}\, .
\end{equation}
Expanding again the action \eqref{eq_coadj_flat} now leads to 
\begin{equation}
\begin{split}
S = -\int \frac{dt dx}{4\pi} \sum_{p_y} &\Biggl[\sum_{q_y}\partial_x \phi_{(-q_y,p_y)} (\dot \phi_{(q_y,p_y)} + \epsilon' \partial_x\phi_{(q_y,p_y)}) \\
	&+  \frac{\epsilon''}3\sum_{q_y,q'_y} \partial_x\phi_{(q_y,p_y)}\partial_x\phi_{(q'_y,p_y)}\partial_x\phi_{(-q_y-q'_y,p_y)}\Biggr] + \cdots
\end{split}
\end{equation}
Taking the continuum limit $N \to \infty$ and Fourier transforming $\tilde \phi_{p_y}(t,x,y) \equiv \int_{-\Lambda}^\Lambda \frac{dq_y}{2\pi}\phi_{(q_y,p_y)}(t,x)$ (the tilde on $\tilde\phi$ serves to remind us that it only contains momentum modes $|q|<\Lambda$), this becomes
\begin{equation}
S
	= -\int \frac{dtdx dy}{4\pi} \sum_{p_y} 
	\left(\partial_x \tilde\phi_{p_y} (\dot{\tilde\phi}_{p_y} + \epsilon'\partial_x \tilde\phi_{p_y}) + \frac{\epsilon''}{3}(\partial_x \tilde\phi_{p_y})^3 + \cdots \right)\, , 
\end{equation}
which corresponds to the action conventionally used in higher-dimensional bosonization \cite{Haldane:1994,PhysRevB.48.7790,PhysRevB.49.10877,PhysRevB.52.4833,Houghton:2000bn} (although the irrelevant corrections $\epsilon'',\epsilon'''$ are usually not treated in that approach). A disadvantage of this abelian description is that a smaller set of Fermi liquid operators are representable as local operators: for example, local operators cannot be resolved beyond the artificial scale $2\pi/\Lambda$.

\section{General 2d Fermi Surfaces and Magnetic Coordinates}\label{sec_mag}

In principle, the approach laid out in Sec.~\ref{sec_flat} can be applied to general Fermi surfaces; here we consider Fermi surfaces of arbitrary shape, in $d=2$ spatial dimensions. A semiclassical kinetic theory approach suggests the dynamics of a Fermi surface can be parametrized by a function $p_F(t,\vb x,\theta)$ or $\phi(t,\vb x,\theta)$ that depends on spacetime but also partly on momentum space. At the quantum level, this object should be viewed as a matrix with indices in the non-commutative phase space direction $(\theta, x_s)$ with $x_s \propto \partial_\theta \vb p_F(\theta)\cdot \vb x$ the component of $\vb x$ parallel to the FS (see Fig.~\ref{fig_FS}). While this matrix seems to have $\sim N^2$ entries if the phase space along the Fermi surface is discretized to contain $N$ points, due to the strongly coupled dynamics of the WZW model the effective degrees of freedom is reduced to $N$.

In practice, carrying this out for a non-flat Fermi surface is unwieldy, because the phase space $(\vb x,\vb p)$ now no longer factorizes into a dispersive direction that enjoys useful gradient expansion (the $(x,p_x)$ direction in Sec.~\ref{sec_flat}),  and $d-1$ non-dispersive directions that have to be treated exactly in the Moyal algebra (the $(y,p_y)$ direction in Sec.~\ref{sec_flat}). One way to circumvent this issue in $d=2$ is to turn on a small magnetic field $B$ \cite{Ye:2024osp}. The gauge-invariant momentum
\begin{equation}
\vb k = \vb p + \vb A\, ,
\end{equation}
is now non-commutative, but it commutes with the guiding center coordinate:%
	\footnote{All our definitions are gauge invariant. Our convention is $\partial_iA_j - \partial_j A_i = - B \epsilon_{ij}$. In symmetric gauge $\vb A = -\frac12 B \hat z \times  \vb x$, the guiding center can be written $\vb R = \frac1B \hat z \times (\vb p - \vb A)$.}
\begin{equation}
\vb R
	= \vb x + \frac1B \hat z \times \vb k\, , 
\end{equation}
Specifically, viewed as single-body operators, these coordinates satisfy the canonical commutation relations
\begin{equation}\label{eq_comrel_kR}
[k_i, k_j] = iB \epsilon_{ij}\, , \qquad
[R_i, R_j] = -\frac{i}B \epsilon_{ij}\, , \qquad
[k_i, R_j] = 0\, ,
\end{equation}
which imply that the Moyal product factorizes:
\begin{equation}
\begin{split}
f\star g
	&= f \exp \left[\frac{i}2(\stackrel{\leftarrow}{\nabla}_x\cdot \stackrel{\rightarrow}{\nabla}_p - \stackrel{\leftarrow}{\nabla}_x\cdot \stackrel{\rightarrow}{\nabla}_p)\right]g\\
	&= f \exp \left[\frac{i}2 B \stackrel{\leftarrow}{\nabla}_{k}\times \stackrel{\rightarrow}{\nabla}_{k}\right] \exp \left[-\frac{i}2 \frac{1}{B} \stackrel{\leftarrow}{\nabla}_{R}\times \stackrel{\rightarrow}{\nabla}_{R}\right]g\\
	&\equiv f (\star_k) (\star_R) g\, .
\end{split}
\end{equation}
The Lie algebra is therefore a product of two Moyal (or GMP) algebras
\begin{equation}\label{eq_mag_algebra}
f\in 
	w_{\infty}^{(k)} \otimes w_{\infty}^{(R)}\, .
\end{equation}
Crucially, (magnetic) translation symmetry implies that the Hamiltonian and ground state are only nontrivial in the first factor
\begin{align}\label{eq_H_f0_mag}
H(\vb k,\vb R)
	&= \epsilon(\vb k) \otimes \mathds 1\,,\\
f_0(\vb k,\vb R) 
	&= 
	f_0(\vb k) \otimes \mathds 1 \simeq \Theta(\vb k\in {\rm FS})\otimes \mathds 1\,. \label{eq_f0_simeq}
\end{align}
(the ``$\simeq$'' in the last equation will be discussed shortly).
This first factor will be treated similarly to 1d bosonization discussed in Sec.~\ref{ssec_1d}. The second factor is instead similar to the $(y,p_y)$ direction in the flat Fermi surface: it is dispersionless, and Moyal products must be evaluated exactly, to all orders in gradients. This noncommutative structure will be treated as for flat Fermi surfaces by first discretizing. In the present context, this does not require a lattice but simply a finite volume $V = L^2$. Imposing periodic boundary conditions $R_i\sim R_i + L$, the non-commutation of coordinates implies that $\vb R$ must take values in a $N_\Phi$ by $N_\Phi$ lattice, with 
\begin{equation}
N_\Phi 
	= \frac{B V}{2\pi}
\end{equation}
the total number of magnetic fluxes through the system. This discretizes the algebra to $u(N_\Phi)$, so that instead of \eqref{eq_mag_algebra}, we will consider the algebra
\begin{equation}\label{eq_mag_algebra_discrete}
f\in
	w_{\infty}^{(k)} \otimes u(N_\Phi)\, .
\end{equation}
The Moyal algebra is recovered in the thermodynamic limit: $\lim_{N_{\Phi}\to \infty} u(N_\Phi) = w_\infty^{(R)}$.

We now comment on the $\simeq$ in \eqref{eq_f0_simeq}. The ground state of a free Fermi gas in a small magnetic field only becomes a sharp Fermi surface $\Theta(\vb k\in {\rm FS})$ as $B\to 0$. While the slight fuzziness of the appropriate state $f_0(\vb k)$ does not affect the leading order dynamics, it is important to correctly capture $T^2$ corrections to specific heat and $q^2$ (or $\omega^2$) corrections to local observables, even as $B\to 0$. To efficiently capture these corrections, it will be useful to define a modified Wigner function $f$ that instead features a sharp Fermi surface for all $B$. We postpone this to Sec.~\ref{sec_corrections}, and focus on the leading low-energy observables in the present section.

\subsection{Leading order action}

Consider a (single) 2d Fermi surface parametrized by a smooth curve $\vb k_F: S^1 \to \mathbb R^2$. We parametrize it with $0\leq \theta < 2\pi$. Following the discussion that lead to \eqref{eq_Poisson_gauge}, we can take $\phi$ to depend on 
\begin{equation}\label{eq_mag_Poisson_gauge}
\phi(\vb k,\vb R) \to 
	\phi(\theta, \vb R) = 
	\phi(\vb k_F(\theta), \vb R)
	\, .
\end{equation}
We will view $\phi(t,\theta,\vb R)\equiv \phi(t,\theta)_{R_1,R_2}$ as the components of the $N_\Phi\times N_\Phi$ hermitian matrix $\phi(t,\theta)$. The coadjoint orbit action again takes the form
\begin{equation}
S = 
	\int dt \Tr \left[f_0 U^{-1} \left(i \partial_t - \epsilon\right)U\right]\, , 
\end{equation}
with Hamiltonian invariant under (magnetic) translations $\epsilon = \epsilon(\vb k)$. The trace is over the algebra \eqref{eq_mag_algebra_discrete}, and $U = e^{i\phi}$. We will find shortly that, to leading order at small wavectors, this description of a 2d Fermi surface essentially reduces to a 1d $U(N_{\Phi})_1$ chiral WZW model for the unitary matrix $g(t,\theta)=e^{i\phi(t,\theta)}$.

By analogy with 1d bosonization in Sec.~\ref{ssec_1d}, we expect Moyal-in-$k$ corrections to be suppressed; let us start by treating them only in the leading non-trivial expansion:
\begin{equation}
\begin{split}
i[\phi,f_0] 
	&= i \phi 2i \sin \left(\frac{B}2 \stackrel{\leftarrow}{\nabla}_k \times \stackrel{\rightarrow}{\nabla}_k\right)f_0\\
	&= - \phi \left(B \stackrel{\leftarrow}{\nabla}_k \times \stackrel{\rightarrow}{\nabla}_k\right) f_0 + \frac1{3!2^2} \phi \left(B \stackrel{\leftarrow}{\nabla}_k \times \stackrel{\rightarrow}{\nabla}_k\right)^3 f_0 + \cdots \\
	&\simeq -B\,  \partial_\theta \phi(\theta,\vb R) \delta_\theta(\vb k \in {\rm FS})
\end{split}
\end{equation}
where $\delta_\theta(\vb k \in {\rm FS}) \equiv \int_0^{2\pi} d\theta \delta^2(\vb k_F(\theta) - \vb k)$ is a one-dimensional Dirac delta function that fires along the Fermi surface.  Dropping further gradients in $k$ (i.e., approximating from here on $\star_k$ as a regular product), similar steps which led to Eq.~\eqref{eq_f_nonab} now give 
\begin{equation}\label{eq_mag_f}
\begin{split}
f -f_0= g\star f_0 \star g^{-1} - f_0
	&\simeq B\, g i \partial_\theta g^{-1} \delta_\theta(\vb k \in {\rm FS})\, .
\end{split}
\end{equation}
One can perform a similar calculation for the energy density
\begin{equation}
g^{-1}\star \epsilon \star g - \epsilon
	= \omega_c(\theta) g^{-1}i \partial_\theta g\, , \qquad
\omega_c(\theta) \equiv \frac{B v_F(\theta)}{|\partial_\theta \vb k_F(\theta)|}\, ,
\end{equation}
where we identified the cyclotron frequency, which here varies along the Fermi surface. The Fermi velocity is defined as the gradient of the dispersion perpendicular to the Fermi surface: $\nabla_k \epsilon(\vb k) = v_F(\theta) \hat z \times \frac{\partial_\theta \vb k_F}{|\partial_\theta \vb k_F|}$. Further following the derivation in Sec.~\ref{ssec_coadj_to_WZW_flat} leads to the action of the chiral $U(N_\Phi)_1$ WZW model, with a position-dependent ``velocity'' $\omega_c(\theta)$:%
	\footnote{This description is similar to that of the edge dynamics of a droplet of quantum Hall ferromagnet in the lowest Landau level \cite{Sakita:1996ne,Ray:2001sz}, with the trapping potential $V(\vb x)$ playing the role of our dispersion relation $\epsilon(\vb k)$. One key difference is that in the present construction, the $SU(N)$ has a spatial interpretation.}
\begin{equation}\label{eq_mag_WZW}
S = -\int \frac{dt d\theta}{4\pi} {\rm tr} \left(\partial_\theta g^{-1} (\partial_t + \omega_c(\theta) \partial_\theta)g\right) - \frac1{12\pi} \int {\rm tr}(g^{-1}dg)^3\, .
\end{equation}
\revchange{We have considered for a Fermi surface of arbitrary shape, but with the topology of a circle for simplicity. The generalization to several connected components is straightforward and would consist of a sum of actions of the form \eqref{eq_mag_WZW}. }

As a first check of this description, let us compute the specific heat. Eq.~\eqref{eq_mag_WZW} can be viewed as a 1+1d CFT in a background metric $ds^2 = -dt^2 + \frac{d\theta^2}{\omega_c(\theta)^2}$. The conformal map to the thermal cylinder produces a thermal expectation value of the stress tensor $\langle T_{\mu\nu}\rangle = \frac{\pi c}{6} \left(2\delta_\mu^0 \delta_\nu^0 + g_{\mu\nu}\right)$. The specific heat is therefore
\begin{equation}
C_V = \frac{dE }{dT}
	= \frac{d}{dT} \int_0^{2\pi} d\theta \sqrt{g} \langle T_{00}\rangle
	= \frac{\pi}{3} c T \int_0^{2\pi} \frac{d\theta}{\omega_c(\theta)}\, .
\end{equation}
Since the central charge of the chiral $U(N_{\Phi})_1$ WZW model is $c = \frac12 N_{\Phi} = \frac12 \frac{BV}{2\pi}$, this becomes
\begin{equation}
C_V
	= V \frac{\pi}{6} T \int_0^{2\pi} \frac{d\theta}{2\pi}\frac{|\partial_\theta \vb k_F(\theta)|}{v_F(\theta)}\, .
\end{equation}
We recognize the density of single particle states at the Fermi surface in the last factor --- this expression indeed reproduces the leading specific heat at low temperatures of a Fermi surface of arbitrary shape. We emphasize that no patches were needed in this derivation: the Fermi surface remains smooth at every intermediate step. This leading order specific heat was obtained from previous approaches to bosonization, either from a patch prescription \cite{PhysRevB.48.7790,PhysRevB.49.10877,Houghton:2000bn} or in an abelian Landau-level bosonization type description \cite{Ye:2024osp} in terms of $N_{\Phi}$ weakly coupled bosons. As we will discuss in the next section, both of these approaches can be obtained as different abelianizations of our description. An advantage of our approach is that it systematically captures corrections to leading order response; these effects will be studied in Sec.~\ref{sec_corrections}.

\subsection{Abelianization}\label{ssec_abelianization_mag}

The $U(N_{\Phi})_1$ WZW model description of a Fermi surface enjoys a vertex representation in terms of $N_{\Phi}$ weakly coupled compact bosons. As for flat Fermi surfaces, this abelianization process depends on a choice of a $U(1)^{N_\Phi}$ subgroup, see Sec.~\ref{ssec_abelianization}. The form of the abelianized action is insensitive to the choice: considering generators $T_m\in u(N_\Phi)$, $m=1,\ldots, N$ normalized as $\tr T_m T_{m'} = \delta_{mm'}$ and following the steps of Sec.~\ref{ssec_abelianization} leads to 
\begin{equation}
S = - \sum_{m=1}^{N_\Phi} \int \frac{dt d\theta}{4\pi} \partial_\theta \phi_m (\dot \phi_m + \omega_c(\theta) \partial_\theta \phi_m)\, .
\end{equation}
For an isotropic Fermi surface $\omega_c(\theta)=$ const, this reproduces Landau level bosonization \cite{PhysRevB.55.R7347}, see also \cite{Ye:2024osp}. The choice of abelianization made there consists in noticing that translation generators $T_{\vb q} = e^{i\vb q\times \hat{\vb R}}$ commute if $\vb q\times \vb q' \in 2\pi B \mathbb Z$ (magnetic Brillouin zone). 

Other choices of abelianization are possible---while they do not change the leading order form of the action, they will change the expression for local Fermi liquid operators when expressed in terms of 1+1d CFT operators. We briefly comment on a choice that has similarities with the patch (or pillbox) prescription of \cite{PhysRevB.48.7790,PhysRevB.49.10877,Houghton:2000bn}. We will follow the discussion from Sec.~\ref{ssec_abelianization}, and in particular the identification of a $U(1)^N$ subgroup in Eq.~\eqref{eq_patch_gen}. The role of $x,p$ is currently played by $R_x,R_y$. The analog of the generators in \eqref{eq_patch_gen} can be obtained by Fourier transforming one of the coordinates, say $R_x\to q_x$. An abelian subgroup is then spanned by
\begin{equation}
f(q_x,R_y)\, , \qquad 0 \leq q_x <   \Lambda\, , \qquad R_y \in \frac{\Lambda}{B} \mathbb Z\, .
\end{equation}
We claim that this abelianization is effectively implementing the patch prescription described in Fig.~\ref{sfig_1b} even though---interestingly---the Fermi surface parameter $\theta$ is kept smooth in our approach. Indeed, we will see shortly that when evaluating local correlators, $\vec R$ is evaluated at $\vec R = \vec x + \frac1{B} \hat z \times \vb k_F(\theta)$ (see Eq.~\eqref{eq_rhox_expression} below). A discretization $\delta R = \Lambda / B$ thus effectively leads to a discretized angle $\delta \theta = k_F/\Lambda$, as in Fig.~\ref{sfig_1b}.

\subsection{Local observables}

Let us now study local observables, focusing on density correlators. The charge density operator is
\begin{equation}\label{eq_rhox_expression}
\begin{split}
\rho(t,\vb x_0)
	&= \int \frac{d^2\vb xd^2\vb p}{(2\pi)^2} \delta^2(\vb x - \vb x_0) f(t,\vb x,\vb p)\\
	&= \int \frac{d^2\vb kd^2\vb R}{(2\pi)^2} \delta^2(\vb R - \tfrac1B \hat z \times \vb k - \vb x_0) f(t,\vb k,\vb R)\\
	&= \int \frac{d\theta}{2\pi}   \left.j_{\vb R}(t,\theta)\right|_{\vb R=\vb x_0 + \frac1B\hat z\times \vb k_F(\theta)}\, , 
\end{split}
\end{equation}
where we used \eqref{eq_mag_f} and denoted the CFT current by $j_{\vb R}(t,\theta) \equiv \frac1{2\pi} \tr \left(g i \partial_\theta g^{-1} T_{\vb R}\right)$, where $T_{\vb R}$ denotes the corresponding generator in $U(N_{\Phi})$. The Fourier transform of this generator $T_{\vb q}$ corresponds to the 't Hooft basis of $U(N_{\Phi})$, Eq.~\eqref{eq_UN_algebra}, which in the continuum limit satisfies
\begin{equation}
\tr T_{\vb q} T_{\vb q'}
	= N_{\Phi} \delta_{\vb q , -\vb q'} \ \xrightarrow{\ N_{\Phi}\to \infty\ } \ 
	\frac{B}{2\pi} (2\pi)^2 \delta^2(\vb q + \vb q')  \, .
\end{equation}
The Fourier transform of the density has a simple expression in terms of the corresponding current:
%
\begin{equation}\label{eq_rho_mag_leading}
\begin{split}
\rho(t,\vb q)
	&= \int {d\theta}\, e^{-i(\vb q \times \vb k_F(\theta))/B}  j_{\vb q}(t,\theta)\, .
\end{split}
\end{equation}
We are now ready to evaluate the density two-point function:
\begin{equation}\label{eq_rho_mag_leading_step}
\begin{split}
\langle \rho(t,\vb q) \rho(0,\vb q')\rangle
	&= \int {d\theta d\theta'} e^{-i(\vb q \times \vb k_F(\theta) + \vb q' \times \vb k_F(\theta'))/B} \langle j_{\vb q}(t,\theta) j_{\vb q'}(0,\theta') \rangle\\
	&\simeq \frac{B}{2\pi} (2\pi)^2\delta^2(\vb q+\vb q')\int \frac{d\theta d\theta'}{(2\pi)^2} \frac{e^{-i\vb q \times(\vb k_F(\theta)-\vb k_F(\theta'))/B}}{[\omega_c(\bar \theta) t - (\theta-\theta')]^2}\, ,
\end{split}
\end{equation}
where we in the second line we evaluated the two-point function of currents $\langle j_{\vb q}(t,\theta)j_{\vb q'}\rangle  \simeq \frac1{(2\pi)^2}\frac{\tr T_{\vb q}T_{\vb q'}}{(\omega_c t - \theta)^2}$ in the CFT, assuming that the separation in angle $\delta \theta \equiv \theta - \theta'$ is small enough so that the cyclotron frequency does not change appreciably, i.e. $\delta \theta \ll \omega_c(\theta)/\partial_\theta \omega_c(\theta)$, as well as $\delta \theta \ll 2\pi$. These approximations are justified when the external wavevector $\vb q$ is much smaller than the radius of curvature of the Fermi surface. With this approximation, we can expand $\vb k_F(\theta) - \vb k_F(\theta') \simeq \partial_\theta \vb k_F(\theta)\delta \theta$ in the exponent and obtain
\begin{equation}
\begin{split}
\langle \rho \rho\rangle(\omega,\vb q)
	&= \frac{B}{2\pi} \int \frac{d\theta d\delta\theta}{(2\pi)^2} dt \frac{e^{-i\vb q \times(\partial_\theta \vb k_F(\theta)\delta\theta)/B + i\omega t}}{(\omega_c(\theta) t - \delta\theta)^2}\\
	&= i\int_0^{2\pi} \frac{d\theta}{(2\pi)^2} \frac{|\partial_\theta \vb k_F(\theta)|}{v_F(\theta)}\frac{\vb q \cdot \hat n(\theta) v_F(\theta)}{\omega -\vb q \cdot \hat n(\theta) v_F(\theta)}\\
\end{split}
\end{equation}
where $\hat n(\theta) = -\hat z \times \partial_\theta \vb k_F(\theta)/|\partial_\theta \vb k_F(\theta)|$ is the unit vector pointing outside the Fermi surface.
This result agrees with the Lindhard continuum for an arbitrary Fermi surface. For a circle, $|\partial_\theta\vb k_F(\theta)| = k_F$ and this reduces to
\begin{equation}\label{eq_rhorho_leading}
\langle \rho\rho\rangle(\omega,q)
	 = \frac{ik_F}{2\pi v_F} \left[-1 + \frac{|s|}{\sqrt{s^2-1}}\right] \, , \qquad \quad s = \frac{\omega}{v_F|\vb q|}\,.
\end{equation}
%

\section{Power-law corrections to 2d Fermi Surface Dynamics from Bosonization}\label{sec_corrections}

We now turn our formalism into a systematic low-energy expansion for the dynamics of two-dimensional Fermi surfaces, focusing on circular Fermi surfaces for simplicity. The goal is to recover power-law corrections to the low-temperature specific heat, as well as the density two-point function $\langle \rho(t,\vb x)\rho(0,0)\rangle$. This generalizes similar ``beyond Luttinger'' corrections that arise in 1d, due to nonlinearities in the fermion dispersion. In higher dimensions, additional unavoidable corrections also arise from the curvature of the Fermi surface.

\subsection{Modified Wigner function for magnetic coordinates}

To efficiently capture power-law corrections to observables, it is useful to slightly modify our definition of the Wigner function $f(\vb k, \vb R)$. The aspect we want to improve on is the fact that $f_0$ does not feature a sharp Fermi surface in \eqref{eq_H_f0_mag}. This can be traced to the fact that in a magnetic field, the single-body Hamiltonian is no longer diagonalized by the plane waves appearing in \eqref{eq_bilinears}, but instead by Landau levels:
\begin{equation}
H_{\rm single}(\vb x, \vb p)
	= \epsilon \left(\frac12(\vb p + \vb A)^2\right)
	= \epsilon(B a^\dagger a) \, ,
\end{equation}
where \eqref{eq_comrel_kR} implies that $a^\dagger=\frac{i}{2}(k_x - i k_y)\sqrt{2/B}$ and $a$ are raising and lowering operators $[a,a^\dagger]=1$, as are $b^\dagger = i(R_x + i R_y)\sqrt{B/2}$ and $b$; see, e.g., Ref.~\cite{ArovasQHE} for a review. We are considering a general isotropic dispersion relation. The second quantized Hamiltonian is
\begin{equation}\label{eq_H_LL_manybody}
\hat H 
	= \int \frac{d^2p}{(2\pi)^2} \epsilon \left(\frac12 (\vb p+\vb A)^2\right) \psi^\dagger_{\vb p}\psi_{\vb p} 
	= \sum_{n\geq 0} \sum_{m = 1}^{N_{\Phi}} \epsilon(Bn) \psi^\dagger_{nm}\psi_{nm}\, ,
\end{equation}
where $\psi_{\vb p} = \sum_{nm} \langle \vb p| nm\rangle \psi_{nm}$ involves the single-particle Landau level wavefunctions $\langle \vb p| nm\rangle$, reviewed in App.~\ref{app_LL}. Because we are interested in introducing a small magnetic field $B \ll k_F^2$ mostly for the purposes of providing an IR regulator, we consider very large fillings
\begin{equation}\label{eq_filling}
1\ll \nu_F \equiv \frac{Q}{N_{\Phi}} = \frac{2\pi n}{B} = \frac{k_F^2}{2B}\, .
\end{equation}
In this limit, the lower bound $n\geq 0$ of the sum over in \eqref{eq_H_LL_manybody} can be ignored (up to an exponentially small error in $\nu_F$). Eq.~\eqref{eq_H_LL_manybody} can then be viewed as $N_{\Phi}$ 1d chiral fermions propagating with momentum $n$, and dispersion $\epsilon (Bn)$, and can therefore straightforwardly be bosonized---this is the Landau level bosonization of Ref.~\cite{PhysRevB.55.R7347}. We will extend this approach in several ways: first, we will consider the nonabelian bosonization of Landau levels, which will allow us to represent more operators locally. Second, we consider arbitrary (isotropic) dispersion relations. Finally, we will formulate a low-energy expansion to systematically capture observables beyond leading order. 

Fermion bilinears have simpler properties if they are defined directly in terms of the operators $\psi_{nm}$ diagonalizing the Hamiltonian. Consider
\begin{equation}\label{eq_ftilde_def}
\tilde f_{m_1m_2}(\theta,n)
	= \sum_{\Delta n} e^{i\theta \Delta n} \psi^\dagger_{n-\frac{\Delta n}{2},m_1}
	\psi_{n+\frac{\Delta n}{2},m_2}\, , 
\end{equation}
with $n\in \frac12 \mathbb Z$ the average Landau level of the particle-hole pair. When $n$ is (half-)integer, the sum runs over even (odd) $\Delta n$. $\tilde f_{m_1m_2}(\theta,n)$ and $f(\vb k,\vb R)$ can be related by using the Landau level wavefunctions, see App.~\ref{app_LL}. The advantage of $\tilde f$ is that it has a sharp ``Fermi-surface'':
\begin{equation}
\langle {\rm FS}|\tilde f_{m_1m_2}(\theta,n) |{\rm FS}\rangle 
	= \delta_{m_1 m_2} \Theta(n \leq \nu_F)\, ,
\end{equation}
with filling $\nu_F$ \eqref{eq_filling}. Furthermore, its algebra is simple. At large filling (or small $B$), one can take the Landau level index $n$ to be continuous, and the commutator of $\tilde f(\theta,n)$ with $\hat H =\sum_n \int \frac{d\theta}{2\pi} H(\theta,n) f(\theta,n)$ is (see App.~\ref{app_LL} for the derivation)
\begin{equation}\label{eq_mag_moyal}
[\hat H, \tilde f(\theta,n)] \xrightarrow{\ B\to 0 \ }
	-H(\theta,n) 2i \sin \frac12 \left({\stackrel{\leftarrow}{\partial}_\theta \stackrel{\rightarrow}{\partial}_n -\stackrel{\leftarrow}{\partial}_n \stackrel{\rightarrow}{\partial}_\theta}\right) \tilde f(\theta,n)\, .
\end{equation}
In this limit, $\tilde f$ is again an element of the algebra $w_\infty \otimes u(N_{\rm \Phi})$: it can be expanded as in \eqref{eq_f_nonab} in terms of CFT operators:
\begin{equation}\label{eq_ftilde_mag_exp}
\begin{split}
\tilde f(\theta,n)
	&= \tilde f_0(n) + \delta(n-\nu_F) \left[i g \partial_\theta g^{-1}\right] 
	+ \delta'(n-\nu_F) \left[-\frac12 \partial_\theta g \partial_\theta g^{-1}\right] + \cdots\, .
\end{split}
\end{equation}
%

\subsection{Bosonized action and specific heat}\label{ssec_S_mag_full}

The action is again given by
\begin{equation}
S = \int dt \Tr \left[\tilde f_0 U^\dagger\left(i \partial_t - \epsilon\right)U\right]
\end{equation}
with $\tilde f_0(\theta,n) = \Theta (\nu_F - n)$, and $\epsilon  =\epsilon(n)$. Because the algebra \eqref{eq_mag_moyal} is exactly the 1d Moyal algebra, the expansion of the action entirely parallels that of flat Fermi surfaces in Sec.~\ref{sec_flat}, and we obtain again
\begin{equation}\label{eq_mag_WZW_full}
S = -\frac1{4\pi} \int \tr \partial_t g^{-1}\partial_\theta g  - \frac1{12\pi}\int \tr (g^{-1}dg)^3 - \frac1{2\pi}\int \tr \left(g^{-1}h(-i\partial_\theta)g\right)\, , 
\end{equation}
with now $h'(n) = \epsilon(n)$. This is a nonabelian bosonized description of Landau level fermions with arbitrary dispersion relation. At leading order in derivatives, one can expand $h(-i \partial_\theta) \simeq \frac12\omega_c(-i\partial_\theta)^2 + \cdots$ (where we identified $\partial_n \epsilon = \frac{B}{k_F}\partial_{k_F}\epsilon = \frac{Bv_F}{k_F}$ with the cyclotron frequency), and recover the action \eqref{eq_mag_WZW}. Eq.~\eqref{eq_mag_WZW_full} furthermore provides the tower of irrelevant corrections to the $U(N_{\Phi})_1$ WZW model that exactly corresponds to 2d free fermions with arbitrary dispersion relation in a magnetic field.

Given that this description exactly reproduces the low energy spectrum of the model up to energies $E_F$ required for a particle-hole pair to reach the lowest Landau level, the specific heat is also exactly reproduced up to exponentially small corrections $\sim e^{-\nu_F} \sim e^{-k_F^2/B}$. However, we still go through the motions of computing the leading low-temperature ($T\ll E_F$) correction to the specific heat: for interacting Fermi liquids, this low-temperature expansion will be a necessary control parameter, and we anticipate that Fermi-liquid corrections to observables can be treated similarly as outlined below.

The leading specific heat correction can be found using conformal perturbation theory. We expand the action \eqref{eq_mag_WZW_full} 
\begin{equation}
S = S_{\rm CFT} + \delta S_1 + \delta S_2 + \cdots
\end{equation}
in a series of irrelevant corrections $\delta S_l$ of dimension $\Delta = 2+l$ to the 1+1d CFT. The first two, $\delta S_1$ and $\delta S_2$ were already discussed in \eqref{eq_SH_expand_flat}, and are given by
\begin{equation}\label{eq_delta_S}
\begin{split}
\delta S_1 
	&= -\partial_\nu^2\epsilon \int \frac{dt d\theta}{2\pi} \frac{i}{3!} \tr \left( g^{-1} \partial_\theta^3 g\right) \, , \\ 
\delta S_2
	&= -\partial_\nu^3 \epsilon \int \frac{dt d\theta}{2\pi} \frac{1}{4!} \tr \left( g^{-1} \partial_\theta^4 g\right)\, .
\end{split}
\end{equation}
The corrections to the free energy $F = \log Z = \tr e^{-\beta H}$ can be obtained by expanding the Euclidean path integral and evaluating the thermal expectation value of these corrections:
\begin{equation}\label{eq_CV_corrections}
\begin{split}
F &=
 F_{\rm CFT} 
 + \frac12 \langle (\delta S_1)^2\rangle_{\beta} 
 - \langle \delta S_2\rangle_{\beta}  + \cdots\\
 &= V \frac{\pi}{12} \frac{k_F}{v_F}T \left(1 + \# T^2 + \cdots \right)\, ,
\end{split} 
\end{equation}
(note that $\langle \delta S_1\rangle_\beta $ vanishes by symmetry). The specific heat is then $C_V = d (TF)/dT$. The thermal expectation values of $\delta S_i$ can be evaluated using current algebra, or with conventional perturbation theory by abelianizing. 

\subsection{Power-law corrections to local observables}

We now turn to evaluating power-law corrections to local observables, focusing on the density two-point function \eqref{eq_rhorho_leading}. Using the mode expansion for the fields $\psi(\vb x) = \sum_{nm}\langle \vb x | nm\rangle \psi_{nm}$, the density operator is
\begin{equation}
\begin{split}
\rho(\vb x)
	&= \psi^\dagger (\vb x) \psi(\vb x) \\
	&= \sum_{m_{1,2},n_{1,2}} \langle n_1 m_1 | \vb x\rangle \langle\vb x | n_2 m_2\rangle \psi^\dagger_{n_1 m_1} \psi_{n_2 m_2}\\
	&\simeq  \int \frac{dn d\Delta nd\theta }{2\pi} \sum_{m_1m_2} \langle n - \tfrac{\Delta n}2,m_1 | \vb x\rangle \langle\vb x | n + \tfrac{\Delta n}2, m_2\rangle e^{-i\Delta n \theta} \tilde f_{m_1m_2}(\theta,n)
\end{split}
\end{equation}
where in the last line we took the continuum limit for $n_1,\,n_2$, and changed variables to $n=\frac{n_1+n_2}2$ and $\Delta n = n_2-n_1$. 
Fourier transforming $\int d^2x e^{-i\vb q\cdot \vb x}$, one has
\begin{equation}\label{eq_rhoq}
\rho(\vb q)
	= \int \frac{dn d\Delta n d\theta}{2\pi} \sum_{m_1,m_2} \langle n-\tfrac{\Delta n}2,m_1 |e^{-i\vb q\cdot\hat{\vb x}} | n+\tfrac{\Delta n}2,m_2 \rangle e^{-i\theta \Delta n} \tilde f_{m_1m_2}(\theta,n)\, , 
\end{equation}
where we introduced the single-body operator $\hat{\vb x} = \vb R  - \frac1{B} \hat{z} \times \vb k$. The $n$ and $m$ sectors factorize, so that
\begin{equation}\label{eq_rho_factors}
\langle n-\tfrac{\Delta n}2,m_1 |e^{-i\vb q\cdot\hat{\vb x}} | n+\tfrac{\Delta n}2,m_2\rangle
	= 
	\langle n-\tfrac{\Delta n}2 |e^{-i\vb q\times \vb k / B} | n+\tfrac{\Delta n}2\rangle 
	\langle m_1 |e^{-i\vb q\cdot \vb R} | m_2 \rangle \, .
\end{equation}
The second factor is simply the 't Hooft basis generator \eqref{eq_UN_algebra} of $U(N_{\Phi})$:
\begin{equation}
\langle m_1 |e^{-i\vb q\cdot \vb R} | m_2 \rangle 
	= (T_{\vb q})_{m_1m_2}\, .
\end{equation}
Indeed, it is straightforward to show using the Baker-Campbell-Hausdorff formula that these matrix elements satisfy $[T_{\vb q},T_{\vb q'}] = 2i\sin \frac{\vb q \times \vb q'}{2B} T_{\vb q + \vb q'}$ and $\Tr T_{\vb q} T_{\vb q'} = N_{\Phi} \delta_{\vb q, - \vb q'}$. The first factor \eqref{eq_rho_factors} corresponds to the dispersive direction: it can be approximated in the regime of interest with a semiclassical expansion (see App.~\ref{app_LL})
\begin{equation}\label{eq_semiclass_leading}
\int {d\Delta n}  \langle \nu_F-\tfrac{\Delta n}2 |e^{-i\vb q\times \vb k/B} | \nu_F+\tfrac{\Delta n}2\rangle e^{-i\theta \Delta n} 
    \simeq  e^{-i\vb q\times \vb k_F(\theta) /B}\, .
\end{equation}
Finally, using the leading expression for $\tilde f$ from \eqref{eq_ftilde_mag_exp}, we recover the expression \eqref{eq_rho_mag_leading} for the density operator
\begin{equation}
\rho(t,\vb q) = \int {d\theta}e^{-i\vb q\times \vb k_F(\theta) /B} j_{\vb q}(t,\theta) + \cdots\, , \qquad
	j_{\vb q}(t,\theta) = \frac{1}{{2\pi}} \tr \left(T_{\vb q} g i \partial_\theta g^{-1}\right) \,.
\end{equation}
Subleading corrections to the semiclassical approximation \eqref{eq_semiclass_leading} and to the operator \eqref{eq_ftilde_mag_exp} will lead to corrections to density response, even for a parabolic band $\epsilon(k) \propto k^2$. These ``geometric'' corrections to local response are studied in App.~\ref{app_LL}. Here, we will focus on the corrections coming from the irrelevant terms in the CFT \eqref{eq_delta_S}, illustrated in the first line of Fig.~\ref{fig_loop_correction}. These can be treated very similarly to the specific heat in \eqref{eq_CV_corrections}:
\begin{equation}
\delta \langle \rho \rho\rangle
	= - \frac{1}{2}\langle \rho (\delta S_1)^2 \rho \rangle
	+ i \langle \rho (\delta S_2) \rho \rangle + \cdots \, .
\end{equation}
There are several ways to evaluate these correlators. One approach that makes the parallel with ``beyond Luttinger'' corrections to 1d bosonization \cite{pereira2007dynamical,RevModPhys.84.1253} most manifest is to abelianize, following one of the choices in \ref{ssec_abelianization_mag}. 
Here we will take $g= e^{i \Sigma'_{\vb q} T_{\vb q}\phi_{\vb q}}$, with the sum $\Sigma'_{\vb q}$ running over a magnetic Brillouin zone. The leading order current has a simple expression in terms of the abelian compact bosons:
\begin{equation}\label{eq_j_ab}
j_{\vb q}(t,\theta) = \frac{N_{\Phi}}{2\pi} \partial_\theta \phi_{\vb q}\, .
\end{equation}
Furthermore, using $\Tr T_{\vb q_1} \cdots T_{\vb q_n} = N_\Phi \delta_{\Sigma_i \vb q_i,0}$, the action and leading irrelevant terms become
\begin{equation}\label{eq_FFG}
\begin{split}
S	
	= - N_{\Phi} \int \frac{dt d\theta}{2\pi} \frac12 \sum_{\vb q}{}' \partial_\theta \phi_{-\vb q} (\partial_t+\omega_c \partial_\theta)\phi_{\vb q} 
	&+ \frac{\partial_\nu^2\epsilon}{3!} \sum_{\vb q\vb q'}{}' \partial_\theta \phi_{\vb q}\partial_\theta \phi_{\vb q'}\partial_\theta \phi_{-\vb q-\vb q'}\\
	&+ \frac{\partial_\nu^3\epsilon}{4!} \sum_{\vb q}{}' \partial_\theta^2 \phi_{\vb q}\partial_\theta^2 \phi_{-\vb q} + \cdots\, .
\end{split}
\end{equation}
There is also a quartic term proportional to $\partial_\nu^3\epsilon$ (see Eq.~\eqref{eq_abelian_perturbation}) which will not contribute at leading order because it is normal ordered.

\begin{figure}
\begin{center}
\begin{tikzpicture}[line width=1pt, scale=0.8]

\def\xsep{4}

\node at (-1.3,0) {$\langle \rho\rho\rangle \;=$};

  \draw (-0.2,0) -- (2.2,0);

\node at (\xsep*0.72,0) {$+$};

\begin{scope}[xshift=\xsep cm]
  \draw (-0.2,0) -- (0.6,0);
  \draw (1.4,0) -- (2.2,0);
  \draw (1.0,0) circle (0.4);
\end{scope}

\node at (1.75*\xsep,0) {$+$};

\begin{scope}[xshift=2*\xsep cm]
  \draw (-0.2,0) -- (2.2,0);
  \draw (1.0-0.18,0-0.18) -- (1.0+0.18,0+0.18);
  \draw (1.0-0.18,0+0.18) -- (1.0+0.18,0-0.18);
\end{scope}

\begin{scope}[yshift=-0.4*\xsep cm]

\node at (-1.3,0) {$\phantom{\langle \rho\rho\rangle} \;+$};

  \draw (1.4,0) -- (2.2,0);
    \draw[distance=2pt] (-0.2,0) arc (180:0:0.8 and 0.4);
    \draw[distance=2pt] (-0.2,0) arc (180:0:0.8 and -0.4);

\node at (\xsep*0.72,0) {$+$};

\begin{scope}[xshift=\xsep cm]
    \draw[distance=2pt] (-0.2,0) arc (180:0:1.2 and 0.4);
    \draw[distance=2pt] (-0.2,0) arc (180:0:1.2 and -0.4);
\end{scope}

\node at (1.75*\xsep,0) {$+$};

\begin{scope}[xshift=2*\xsep cm]
  \draw (-0.2,0) -- (2.2,0);
  \draw (-0.2-0.18,0-0.18) -- (-0.2+0.18,0+0.18);
  \draw (-0.2-0.18,0+0.18) -- (-0.2+0.18,0-0.18);
\end{scope}

\node at (2.9*\xsep,0) {$+ \quad \cdots $};

\end{scope}

\end{tikzpicture}

\caption{\label{fig_loop_correction}Power-law corrections to local density response can be obtained similarly to 1d bosonization. The diagrams in the first line come from corrections to the dispersion relation and have direct analogs in 1d. Those in the second line do not: they arise from the geometry of the Fermi surface.}
\end{center}
\end{figure} 

Let us first determine how these terms affect the two-point function of the CFT current \eqref{eq_j_ab}, which we will Fourier transform $\langle j_{\vb q}j_{\vb q'}\rangle(\omega,\ell) \equiv \int dt d\theta\, e^{i\omega t - i\ell\theta} \langle j_{\vb q}(t,\theta) j_{\vb q'}(0,0)\rangle$. It will receive a one-loop correction involving two $\partial_\nu^2 \epsilon$ vertices, and a tree-level correction from the $\partial_\nu^3 \epsilon$ vertex (see Fig.~\ref{fig_loop_correction}). These evaluate to
\begin{equation}
\begin{split}
\langle j_{\vb q} j_{\vb q'}\rangle(\omega,\ell)
	&= \frac{i}{2\pi} \frac{N_{\Phi} \delta_{\vb q + \vb q' ,0}}{\omega_c} \frac{\omega_c \ell}{\omega - \omega_c \ell} \\
	&\times \left[
	1 + \frac1{12} \left(\frac{\partial_\nu^2 \epsilon}{\omega_c}\right)^2 \ell^2 \left(\frac{\omega_c\ell}{\omega - \omega_c\ell}\right)^2 + \frac1{12} \frac{ \partial_\nu^3\epsilon}{\omega_c} \ell^2 \frac{\omega_c \ell}{\omega - \omega_c \ell} + \cdots \right]\, ,
\end{split}
\end{equation}
These corrections to the CFT current two-point function are in fact essentially identical to those of 1d bosonization, except with derivatives of the dispersion replaced as $\partial_k^m\epsilon \to \partial_{\nu}^m \epsilon$. The 1-loop correction $\propto (\partial_\nu^2\epsilon)^2$ has an on-shell enhancement near $\omega \approx \omega_c \ell$ familiar from 1d bosonization, signalling the breakdown of (bosonic) perturbation theory and the opening of a particle-hole continuum \cite{RevModPhys.84.1253}. In our higher-dimensional context, this 1d expression will be integrated to give the full density two-point function (see below), which removes the on-shell singularity. Higher dimensional bosonization therefore does not seem to suffer from the on-shell breakdown that occurs in 1d. Following the same steps as in Eq.~\eqref{eq_rho_mag_leading_step}, one finds that the correction to the density two-point function is
\begin{equation}\label{eq_rhorho_corrections}
\begin{split}
\langle \rho\rho\rangle(\omega,q)
	&= \int d\theta \langle j_{\vb q} j_{-\vb q}\rangle(\omega, \ell = \tfrac{k_F q_n}{B})\\
	&= \frac{ip_F}{2\pi v_F} \int \frac{d\theta}{2\pi} \frac{q_n}{\frac{\omega}{v_F }- q_n}
		\left[
	1 + 
	\frac1{12}\left(\frac{q_n}{p_F}\right)^2 \left\{ \frac{p_F^6\epsilon''^2}{v_F^2}\left(\frac{q_n}{\frac{\omega}{v_F} - q_n}\right)^2 
	+ \frac{p_F^5\epsilon'''}{v_F} \frac{q_n}{\frac{\omega}{v_F} - q_n} \right\}+ \cdots \right]
\end{split}
\end{equation}
where $q_n\equiv  \vb q\cdot \hat n(\theta) = \vb q \cdot \vb k_F(\theta)/k_F$, and $(\cdot)'$ denotes derivatives with respect to $k_F^2/2$. These agree with the subleading corrections to the Lindhard function, see App.~\ref{app_FFG}.

\revchange{
Two expansions led to \eqref{eq_rhorho_corrections}: large filling $\nu_F\gg 1$ \eqref{eq_filling}, and small wavevector $q\ll k_F$. The former corresponds to small field $B$ in units of chemical potential; when $B\to 0$,  only the leading terms in this expansion contribute. The latter is the long wavelength approximation of a Fermi liquid, with our calculation capturing the leading and subleading terms in this expansion. Note that the total magnetic flux $N_{\Phi} = BV / (2\pi)$ was treated exactly throughout our calculation: while we have in mind taking the thermodynamic limit first, so that $N_{\Phi}\to \infty$, this is not necessary in practice.}

\section{Discussion}\label{sec_discussion}

We have found that certain nonlinear terms in the bosonized description of Fermi liquids \cite{Delacretaz:2022ocm}, necessary to capture the nonlinear response that is inevitable in $d>1$ dimensions, are in fact relevant and cannot be expanded perturbatively: they lead to strongly coupled---but solvable---dynamics. This in particular resolves the naive overcounting of degrees of freedom when particle-hole pairs are viewed as the fundamental excitation; it gives a concrete way to make sense of fields depending on phase space, which arise in the bosonization of Fermi surfaces \cite{PhysRevB.19.320,Haldane:1994,PhysRevB.48.7790,PhysRevB.49.10877,PhysRevB.52.4833,Houghton:2000bn} or in semiclassical kinetic theory, as genuine quantum fields.
More specifically, we have shown that general smooth Fermi surfaces in $d>1$ are captured by a particular $N\to \infty$ limit of the 1+1d $U(N)_1$ WZW model with a tower of irrelevant corrections. This approach furthermore systematically captures Fermi surface dynamics beyond the leading order low-energy response, extending well-known ``beyond Luttinger'' corrections in 1d bosonization \cite{pereira2007dynamical,RevModPhys.84.1253} to higher dimensions.

This approach reveals connections between Fermi liquids, non-commutative geometry, and 1+1d CFTs. Connections between 1+1d CFTs and higher dimensional Fermi surfaces had been suspected before, partly due to similarities in their entanglement structure \cite{Swingle:2009bf,Swingle:2010yi}; we hope that this recasting of Fermi liquids in the framework of 1+1d CFTs, where many nonperturbative tools are available, will help make progress in the study of compressible phases. Our exact treatment of the noncommutative nature of phase space may also help sharpen notions of symmetries and anomalies of Fermi surfaces \cite{Else:2020jln,Lee:2022hcm,Lu:2023emm}.

While we have focused on the free Fermi gas, where an exact bosonization duality could be established, the central motivation for bosonization is to be able to treat strongly coupled Fermi liquids, and possibly non-Fermi liquids,%
	\footnote{Our finding that Fermi liquids are described by nonlinear sigma models resonates with the observation that perturbative corrections in NFLs are large-$N$ matrix-like rather than large-$N$ vector-like \cite{Lee:2009epi}.}
in terms of weakly coupled bosonic degrees of freedom. We expect our formalism to be particularly useful in this context: we further comment on these future directions below.

\paragraph{Landau parameters:} For an arbitrary Fermi surface, Landau parameters can be introduced by adding a term to the action \cite{Delacretaz:2022ocm}
\begin{equation}
S_{\rm int}
	= \int \frac{dt d^2 \vb xd^2 \vb kd^2 \vb k'}{(2\pi)^2} \delta f(\vb x,\vb k)\delta f(\vb x,\vb k') F_{\vb k,\vb k'}\, .
\end{equation}
We start by focusing on the leading order in $q$ effects of this term, in which case the approach of Sec.~\ref{sec_mag} can be used. Changing variables to $\vb R = \vb x + \frac1{B}\hat z\times \vb k$, this becomes:
\begin{align}
S_{\rm int}
	&= \int_{t\vb R \vb k\vb k'} \delta f(\vb k,\vb R) \delta f(\vb k',\vb R + \tfrac1B \hat z \times (\vb k' - \vb k)) F_{\vb k, \vb k'}\\
	&\simeq \int dt d^2 \vb R d\theta d\theta'  j_{\vb R} (\theta) j_{\vb R'(\theta,\theta')}(\theta') F_{\vb k_F(\theta), \vb k_F(\theta')}\, , \qquad \vb R'(\theta,\theta') = \vb R + \tfrac1B \hat z\times  (\vb k_F(\theta') - \vb k_F(\theta))\, .\notag \\
	&= \int dt d^2\vb{q} d\theta d\theta' e^{-i\vb{q}\times (\vb{k_F}(\theta) - \vb{k_F}(\theta'))/B} j_{\vb{q}}(t, \theta) j_{-\vb{q}}(t, \theta') F_{\vb{k}_F(\theta), \vb{k}_F(\theta')}\, , \notag
\end{align}
where in the last line we Fourier transform to $j_{\vb{q}}$ with  
\begin{equation}
j_{\vb{q}} = \frac{1}{2\pi}\tr(gi\partial_\theta g^{-1} T_{\vb{q}}) = \frac{1}{2\pi}e^{i q_x q_y/2B}\sum_{m=0}^{N_\Phi -1} j_{m,m-n_2}(\theta) e^{i m \frac{q_x L}{N_\Phi}}\, ,
\end{equation}
with $n_{q_y} = \frac{q_y L}{2\pi} \in \{0, \cdots, N_\Phi - 1\}$ and $j_{m,m'}(\theta) = (g i \partial_\theta g^{-1})_{mm'}$. We show in App.~\ref{app_LP_corr} that this interaction reproduces the leading correction to density response from Landau parameters. 
Notably, while the free Fermi gas action in abelianized form \eqref{eq_FFG} only involves derivatives of the $N_{\Phi}$ compact bosons, we find that the Landau parameters involve vertex operators in general (off-diagonal components $j_{mm'}$)%
    \footnote{
    Our expression thus differs from previous similar expressions in the literature  \cite{Barci:2018lsp,Wang:2025asx}. Ref.~\cite{Wang:2025asx} studied global observables, which do not seem to be sensitive to this difference.}.

The approach of Sec.~\ref{sec_corrections} allows to systematically improve on this leading approximation to the Landau parameters, similar to the irrelevant corrections arising from the dispersion relation in Eq.~\eqref{eq_delta_S}.  Schematically, this leads to 
\begin{equation}\label{eq_Landau_int}
S_{\rm int} 
	= \sum_{m,m'}^{N_{\Phi}}\int dt d\theta d\theta' \left[\partial_\theta \phi_m + (\partial_\theta \phi_m)^2\partial_\nu+ \cdots\right]_\theta \left[\partial_\theta \phi_{m'} + (\partial_\theta \phi_{m'})^2\partial_\nu+ \cdots\right]_{\theta'} F_{\theta,\theta'}\, , 
\end{equation}
where the derivatives $\partial_\nu$ act on the Landau parameters, which generically depend on density \cite{Delacretaz:2025ifh}. These corrections are interesting, because they lead to qualitatively new non-analyticities in Fermi liquids \cite{PhysRevB.68.155113,PhysRevB.69.121102,PhysRevB.73.045128}. These are further discussed below.

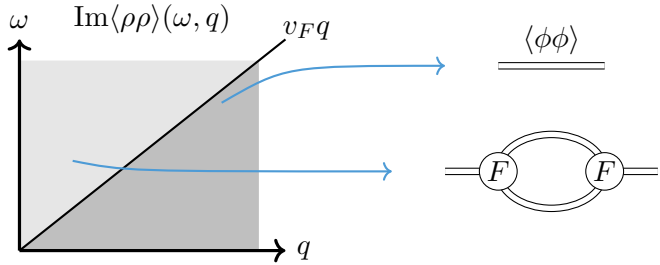
\begin{figure}
\begin{center}
\begin{tikzpicture}[scale=0.7]

\def\vF{0.8} 

\fill[gray!50] (0,0) -- (4.5,0) -- (4.5,{\vF*4.5}) -- cycle;

\fill[gray!20] (0,0) -- (0,{\vF*4.5}) -- (4.5,{\vF*4.5}) -- cycle;

\draw[thick] (0,0) -- (5,{\vF*5}) node[pos=0.95,above right] {$v_F q$};

\node at (2.5,5.5*\vF) {$\Im \langle \rho\rho\rangle(\omega,q)$};

\draw[very thick,->] (0,0) -- (5,0) node[right] {$q$};
\draw[very thick,->] (0,0) -- (0,4) node[above] {$\omega$};

  \begin{scope}[xshift=10cm, yshift=2.5cm]

    \draw[double distance=2pt] (-1,1) -- (1,1);
    \node at (0,1.5) {$\langle \phi\phi\rangle$};

    \draw[double distance=2pt] (-2,-1) -- (-1,-1);
    \draw[double distance=2pt] (1,-1) -- (2,-1);

    \draw[double distance=2pt] (-1,-1) arc (180:0:1cm and 0.7cm);
    \draw[double distance=2pt] (-1,-1) arc (180:0:1cm and -0.7cm);

    \node[draw,circle,fill=white,inner sep=1pt] at (-1,-1) {$F$};
    \node[draw,circle,fill=white,inner sep=1pt] at (1,-1) {$F$};
  \end{scope}

  \draw[->,thick,blue!70] (3.8,2.8) .. controls (5,3.5) .. (8,3.5);

  \draw[->,thick,blue!70] (1,1.7) .. controls (2,1.5) .. (7,1.5);

\end{tikzpicture}
\end{center}
\caption{\label{fig_Nonanalytic} A single bosonic propagator produces the particle-hole continuum $\omega\leq v_F q$, Eq.~\eqref{eq_rhorho_leading}. Landau interactions \eqref{eq_Landau_int} lead to nonzero spectral densities everywhere due to the two-particle-hole continuum.}
\end{figure}

\paragraph{Non-analytic response in Fermi liquids and two-particle-hole continuum:} Interactions in Fermi liquid theory qualitatively change response functions: not only do they allow for novel collective excitations (zero-sound, shear sound, etc.), they also produce corrections with a different analytic structure. Paralleling the multi-particle continuum that interactions produce for regular excitations, Landau parameters lead to a multi-particle-hole continuum above the usual (Lindhard) particle-hole continuum, illustrated in Fig.~\ref{fig_Nonanalytic}. The leading diagram responsible for this continuum is a 1-loop diagram involving cubic vertices from \ref{eq_Landau_int}. In a fermionic description, this would correspond to a 3-loop diagram \cite{PhysRevB.68.155113}.

A simple scaling argument shows that this diagram gives a $O(q^{d+1})$ correction to the density $\rho$ two-point function, and $O(q^{d-1})$ correction to the spin density $s^i$ two-point function. Because these interactions involve different angles, we expect the nonperturbative 1d physics uncovered in this paper not to play a role, and the scaling can be obtained by expanding the semiclassical EFT of \cite{Delacretaz:2022ocm}. The Gaussian part of the action for charge and spin fluctuations is \eqref{eq_S_intro}, implying that $\phi_c\sim \phi_s \sim q^{(d-1)/2}$,  while the leading nonlinearities have the form \cite{Delacretaz:2022ocm}
\begin{equation}
\delta S_3^{\rm charge} \sim \int dt d^dx d^{d-1}\theta \, (\nabla \phi_c)^3\, , \qquad
\delta S_3^{\rm spin} \sim \int dt d^dx d^{d-1}\theta \, \phi(\nabla \phi_s)^2\, .
\end{equation}
The cubic vertices are therefore suppressed by $\delta S_3^{\rm charge}/S_2 \sim \nabla \phi_c \sim q^{(d+1)/2}$ and $\delta S_3^{\rm spin}/S_2 \sim \phi_s \sim q^{(d-1)/2}$. The scaling of the correction follows from using two of these vertices. In summary, we have
\begin{align}
\langle \rho\rho\rangle(\omega,q)
	&= F_0(s) + q^2F_2(s) + q^{d+1}F_{d+1}(s) + \cdots \, , \\
\langle s^z s^z\rangle(\omega,q)
	&= \tilde F_0(s) + q^2\tilde F_2(s) + q^{d-1}\tilde F_{d-1}(s) + \cdots\, ,
\end{align}
with $s=\omega/(v_F q)$. The $q^2$ correction already arises for a free Fermi gas, and was studied in Sec.~\ref{sec_corrections}.
The first $q^{d-1}$ non-analytic correction to the spin two-point function is well-known \cite{PhysRevB.68.155113,PhysRevB.69.121102,PhysRevB.73.045128}, as is the fact that this correction does not enter in the density correlator due to cancellations. Our scaling argument implies that the leading non-analytic correction to density response is $O(q^{d+1})$ which, as far as we know, is a new result. The simple scaling argument we have used is not available in the fermionic description due to approximate cancellations.

\paragraph{Vertex operators and BCS interaction:} Our formulation of higher-dimensional Fermi liquids in terms of a 1+1d CFT allows to carry over certain tools from 1d bosonization. In particular, it is possible to use a vertex representation of the $U(N)_1$ WZW model to represent the fermion as a vertex operator, albeit nonlocally, and to locally represent charge-two fermion bilinears. These are particularly interesting from the perspective of the Fermi liquid EFT \cite{RevModPhys.66.129,Polchinski:1992ed}, since they are responsible for the BCS interaction.

\paragraph{Fermi liquids in weak magnetic fields:} A small $B\ll k_F^2$ was used in Secs.~\ref{sec_mag}, \ref{sec_corrections} as a trick leading to a useful factorization of phase space, with the intention to set $B\to 0$ in the end. However, we anticipate that our approach may be useful to study Fermi liquid response with a small nonzero $B$. Following earlier work on bosonization in a small magnetic field \cite{PhysRevB.55.R7347,Barci:2018lsp}, recent work by Ye and Wang \cite{Ye:2024osp,Ye:2024pty,Wang:2025asx} has elegantly established nonperturbative results on magnetic oscillations in various observables. One contribution of our work in this direction is that it shows how to access local probes, which have a more complex nonlinear structure. It would also be interesting to use our formalism to capture the behavior of collective excitations in a magnetic field \cite{Barci:2018lsp,Nguyen:2018zlb}. 

\paragraph{Other abelianization schemes:} One appeal of the nonabelian description is that it keeps a smooth Fermi surface (Fig.~\ref{fig_FS}) and preserves spatial symmetries. However, as we have seen, abelianizing is useful to replace conformal perturbation theory with regular perturbation, and we expect this may be the simplest approach to study (non-)Fermi liquids. 
In previous approaches to abelian Landau level bosonization \cite{PhysRevB.55.R7347,Ye:2024osp}, a given abelianization is implicitly assumed (corresponding to a choice of a magnetic Brillouin zone in \cite{Ye:2024osp}); other choices however are possible, as discussed in Sec.~\ref{ssec_abelianization} and \ref{ssec_abelianization_mag}. While different choices do not change the abelianized action, they affect the expression for local Fermi liquid operators in terms of operators of the 1+1d CFT. Making an appropriate choice may make more tractable the study of Fermi liquids coupled locally to other degrees of freedom.

\subsection*{Acknowledgments}

We thank Andrey Chubukov, Eduardo Fradkin, Diego Garc\'ia-Sep\'ulveda, Emil Martinec, Dmitrii Maslov, Umang Mehta, Riccardo Rattazzi, Dam Thanh Son, Mike Stone, Yuan Wan, Yuxuan Wang, Paul Wiegmann, Xiaochuan Wu, and Mengxing Ye for many useful discussions. This work was supported by a NSF CAREER award (DMR-2441227).

\appendix

\section{Approaches to bosonization}\label{app_approaches_boso}

\subsection{Bulk vs.~Boundary bosonization}

One-dimensional bosonization is one of the simplest QFT dualities, providing a beautiful example of how degrees of freedom can reorganize in quantum many-body physics. The continuum or thermodynamic limit is essential for such a nontrivial reorganization. 

On a finite lattice, there is still a more straightforward representation of the dynamics of fermions in terms of a fermion bilinear, e.g.~through Hubbard-Stratonovich integration or with a coherent state path integral. In this context, the degree of freedom corresponds to a finite-momentum particle-hole pair $\phi_{k_1 k_2}$ extending into the {\em bulk} of the Fermi sea, see, e.g., Refs.~\cite{Das:1991uta,Dhar:1992rs,Khveshchenko:1993ug,PhysRevB.52.10877,Kavalov:1996ab,Sakita:1996ne,Das:2003vw,Karabali:2003bt,Karabali:2006eg,PhysRevB.74.075102,PhysRevB.82.235120,Park:2023coa}. The action still has the form \eqref{eq_coadj_action}; however, the gauge redundancy due to the stabilizer \eqref{eq_stabilizer} does not allow to remove the $p$ dependence of $\phi$---instead, $\phi_{k_1k_2}$ can at most be reduced to a matrix that has nonzero components for $k_1<k_F$ and $k_2> k_F$, representing a finite particle-hole excitation (see Ref.~\cite{Park:2023coa}).

This ``bulk'' bosonization approach does not seem to offer an advantage compared to working directly with fermions---in particular, in an interacting Fermi (or Luttinger) liquid, these bulk excitations are not weakly coupled. Nevertheless,  our approach for bosonizing higher-dimensional Fermi surfaces in some sense combines aspects of boundary and bulk bosonization. This is most clear in the description of flat Fermi surfaces (Sec.~\ref{sec_flat}): the wire direction is treated by regular 1d (boundary) bosonization, while the other direction is discrete and can taken to be finite ($N$ wires). 

\revchange{
We emphasize that the regular (``boundary'') bosonization of 1+1d free fermions is exact, and can also capture particle-hole excitations  extending into the bulk of the Fermi surface. The expression directly follows from \eqref{eq_bilinears}
\begin{equation}
\psi^\dagger_{k_1}\psi_{k_2}
    = \int dx \, e^{i(k_1 - k_2) x} f(x,p=\tfrac12(k_1+k_2))
\end{equation}
This can also be done in the conventional approach to one-dimensional bosonization (reviewed in App.~\ref{app_oldschool_boso_match}), where this operator is expressed as
\begin{equation}
\psi^\dagger_{k_1}\psi_{k_2}
    = \frac{1}{2\pi} \int dx_1 d x_2 \, e^{i(k_1 x_1 - k_2x_2)} :e^{i(\phi(x_1) - \phi(x_2))}:
\end{equation}
Both expressions are in fact equivalent, as can be shown following the discussion in App.~\ref{app_oldschool_boso_match}. The ``bulk bosonization'' representation of this operator is straightforward, $\phi_{k_1k_2} = \psi^\dagger_{k_1} \psi_{k_2}$.
}

It would be interesting to better understand the connection between these bulk and boundary perspectives (or Fermi sea and Fermi surface), and why the gauge fixing \eqref{eq_Poisson_gauge} allows one to go from bulk to boundary description when $N\to \infty$. See \cite{Sakita:1996ne,Karabali:2003bt,Karabali:2006eg,Polychronakos:2007oda,Cappelli:2018dti,Cappelli:2021kxd} for related discussions in the context of quantum Hall droplets.

\subsection{Lightning review of 1d bosonization}\label{app_oldschool_boso_match}

A right-moving Weyl fermion
\begin{equation}
S = \int dt dx \, \psi^\dagger i(\partial_t + \partial_x) \psi
\end{equation}
has a dual description in terms of a compact chiral boson%
	\footnote{More precisely, fermion parity $(-1)^F$ is gauged in the latter description \cite{Karch:2019lnn}. }
\begin{equation}
S = - \frac1{4\pi}\int dt dx \, \partial_x \phi (\partial_t + \partial_x)\phi\, .
\end{equation}
Both have a global $U(1)$ symmetry with chiral anomaly coefficient $k=1$. The spectrum of local vertex operators is \cite{francesco2012conformal}
\begin{equation}
V_n = e^{-2in\phi}\, , \qquad \bar h = 0,\, \quad h=2n^2\, , 
\end{equation}
which correspond to the even charge fermion bilinears $\psi \partial\psi\ (h=2)$,  $\psi \partial\psi \partial^2 \psi\ (h=8)$, etc. The fermion operator $\psi$ is not locally represented in terms of the boson: while it is tempting to write
\begin{equation}\label{eq_1dboso_psi}
\psi \sim \frac1{\sqrt{2\pi}}e^{-i\phi}\, .
\end{equation}
which has the correct unit $U(1)$ charge, this vertex operator is ill-defined. The appropriate operator is tied to a string (``Klein factor'') and is nonlocal. Nevertheless, this string vanishes when considering neutral fermion bilinears, so that \eqref{eq_1dboso_psi} is useful for this purpose. For example, the density operator can be obtained
\begin{equation}
\rho = \lim_{\delta\to 0}\psi^\dagger(x+\tfrac{\delta}{2})\psi(x-\tfrac{\delta}{2})
	= \frac{1}{2\pi} \partial_x\phi\, ,
\end{equation}
where we used the OPE of two vertex operators and dropped a UV divergent constant.

As a check of our results in Sec.~\ref{ssec_1d}, we will apply the same approach to  obtain a more complicated bilinear: the Hamiltonian density for a right-moving fermion with arbitrary dispersion relation $\epsilon(p_x)$:
\begin{equation}
\mathcal H'
	= \psi^\dagger \epsilon(-i \partial_x) \psi
	= 
	\underbrace{\psi^\dagger \epsilon \left(-i \tfrac12(\stackrel{\rightarrow}\partial_x - \stackrel{\leftarrow}{\partial_x})\right) \psi}_{\mathcal H}
	\, +\,  \hbox{total derivative}.
\end{equation}
We will work with $\mathcal H$. Using \eqref{eq_1dboso_psi} and the vertex operator OPE again, it can be expressed
\begin{equation}\label{eq_Hcal_temp}
2\pi \mathcal H 
	= \lim_{\delta\to 0} \epsilon(-i \partial_\delta) \left[e^{i(\phi(x+\frac{\delta}{2}) - \phi(x-\frac{\delta}{2}))} - 1\right] \left(\frac{-i}{\delta}\right)\, ,
\end{equation}
where the ``-1'' removes the UV divergence in the OPE. Now, use the fact that for two functions $h$ and $g$,
\begin{equation}
(-i)\lim_{\delta\to 0}
	h'(-i \partial_\delta) \frac{g(\delta) - g(0)}{\delta}
	= \lim_{\delta\to 0}h(-i \partial_\delta) g(\delta) - h(0) g(0)\, .
\end{equation}
Returning to \eqref{eq_Hcal_temp} and applying this identity with $g(\delta) = e^{i(\phi(x+\frac{\delta}{2}) - \phi(x-\frac{\delta}{2}))}$ and $h' = \epsilon$, we find:
\begin{equation}
2\pi \mathcal H 
	= -h(0) + \lim_{\delta\to 0}h(-i \partial_\delta) e^{i[\phi(x+\frac{\delta}{2}) - \phi(x-\frac{\delta}{2})]}\, , 
\end{equation}
which agrees with \eqref{eq_SH_1d_neat}.

\subsection{Non-abelian bosonization}\label{app_nonabelianboso}

$N$ Weyl fermions can be bosonized individually following the approach above, which we will refer to as abelian bosonization. Alternatively, as discussed in Sec.~\ref{sec_flat}, they also admit a description that makes the $U(N)$ symmetry manifest, in terms of the $U(N)_1$ WZW model \cite{Witten:1983ar,Polyakov:1984et,Knizhnik:1984nr}.%
	\footnote{Separating the Weyl fermions into two Majorana fermions $\psi = \chi_1 + i \chi_2$, it is possible to make the larger $O(2N)$ symmetry manifest by using the $O(2N)_1$ WZW model. However, working with $U(N)$ is more natural in the context of higher-dimensional Fermi surfaces, due to its connection with the $w_\infty$ algebra and Moyal (or Poisson) brackets.}
In Sec.~\ref{sec_flat}, we used the coadjoint orbit approach to obtain the nonabelian bosonized action for $N$ fermions with arbitrary dispersion $\epsilon(p_x)$. In this appendix, we will use the nonabelian bosonization dictionary to confirm our result. For this purpose, it is useful to consider the non-chiral $U(N)_1$ model with both left and right movers, for which one has the following operator correspondence \cite{Witten:1983ar}
\begin{equation}
\psi_{iR}^\dagger(z) \psi_{jL}(\bar z)
	= i M g_{ij}(z,\bar z)\, ,
\end{equation}
and then focus on purely holomorphic composite operators. In this expression, $M$ is a UV dependent scale that accounts for the anomalous dimension $\Delta_g = 1$. $g$ is often factored in a $SU(N)$ piece and a $U(1)$ piece, however we will find it convenient to keep it as a $U(N)$ matrix. 
Using the fact that $g$ is unitary one has
\begin{equation}
iM (g^{-1})_{ij}
	= iM (g_{ji})^\dagger
	= \psi_{iL}^\dagger (\bar z) \psi_{jR}(z)\, .
\end{equation}
One can thus compute
\begin{equation}
\begin{split}
M^2 
	[g^{-1} \partial^n g]_{ij}
	&= -  \psi_{iL}^\dagger (\bar z) \psi_{kR}(z) \partial^n (\psi_{kR}^\dagger \psi_{jL})\\
	&\supset n\partial G(\bar z)\delta_{kk} \psi_{iR}^\dagger (z)  \partial^{n-1}\psi_{jR}\\
	&=\frac{-i}2 \delta^2(0)N n \psi_{iR}^\dagger (z)  \partial^{n-1}\psi_{jR} \, ,
\end{split}
\end{equation}
where in the last line we used the fact that the fermion Green's function satisfies $\partial G(\bar z) = \frac{-i}{2\pi}\partial \frac{1}{\bar z} = \frac{-i}{2} \delta^2(\vec x)$. In the second line, we only kept terms where the left-moving operators, absent in the chiral model we are considering, fuse to the identity. For $n=1$, matching the current $j = \frac1{2\pi}gi\partial g^{-1} = \psi^\dagger \psi$ fixes 
\begin{equation}
M^2 = \frac{1}{4\pi}\delta^2(0) N\, .
\end{equation}
We can now obtain irrelevant operators:
\begin{equation}
\frac1{2\pi }[g^{-1} (-i\partial)^n g]_{ij} = n\psi^\dagger_i (-i\partial)^{n-1} \psi_j
\end{equation}
so for $\epsilon = h'$,
\begin{equation}
\frac1{2\pi}[g^{-1} h(-i\partial) g]_{ij} = \psi^\dagger_i \epsilon(-i\partial) \psi_j
	\qquad \Rightarrow \qquad
	\frac1{2\pi} \Tr g^{-1} h(-i\partial) g = \psi^\dagger \epsilon(-i\partial ) \psi\, .
\end{equation}
This confirms the expression found in Eq.~\eqref{eq_SH_nonab_again}. Note that for the chiral model, $\bar \partial = 0$ so that $\partial = \partial_x$.

\section{Landau levels and bosonization}\label{app_LL}

\subsection{Magnetic Moyal algebra}

In Sec.~\ref{sec_corrections}, we found it useful to define a modified Wigner function \eqref{eq_ftilde_def}, whose definition we copy here:
\begin{equation}
\tilde f_{m_1m_2}(\theta,n)
	= \sum_{\Delta n} e^{i\theta \Delta n} \psi^\dagger_{n-\frac{\Delta n}{2},m_1}
	\psi_{n+\frac{\Delta n}{2},m_2}\, .
\end{equation}
The Fourier transform of this object,
\begin{equation}
\tilde f(\Delta n,\varphi)
	= \int_0^{2\pi} \frac{d\theta}{2\pi} \sum_{\Delta n} e^{in\varphi - i \Delta n \theta} \tilde f(\theta,n)\, ,
\end{equation}
satisfies the $w_\infty$ algebra
\begin{equation}
[\tilde f(\Delta n, \varphi),\tilde f(\Delta n', \varphi')]
	= 2i \sin \frac12 (\Delta n\varphi' - \varphi\Delta n') \tilde f(\Delta n + \Delta n',\varphi + \varphi')\, .
\end{equation}
We have dropped the dependence on $m_1,\,m_2$, since it is not essential to this discussion. The discreteness of the levels implies that Moyal bracket obtained upon Fourier transforming is a little more complicated (although this would also arise in conventional 1d bosonization in a finite volume): for $\hat H =\sum_n \int \frac{d\theta}{2\pi} H(\theta,n) f(\theta,n)$, 
\begin{equation}
\begin{split}
[\hat H, \tilde f(\theta,n)]
	&= -H(\theta,n) 2i \sin \frac12 \left(-i{\stackrel{\leftarrow}{\partial}_\theta {\rm asin}(i\stackrel{\rightarrow}{d}_n) + i\stackrel{\rightarrow}{\partial}_\theta {\rm asin} (i\stackrel{\leftarrow}{d}_n) }\right) \tilde f(\theta,n)\\
	&\equiv -[H(\theta,n),\tilde f(\theta,n)]_{\rm MagMB} \,  ,
\end{split}
\end{equation}
where $d_n$ is a discrete derivative. The arcsine produces additional higher order in $\partial_\nu \sim B$ corrections that are not accompanied by $\partial_\theta\sim B$. Therefore, they can be ignored in the $B\to 0$ limit, where the Landau levels become continuous and one has 
\begin{equation}
[H(\theta,n),\tilde f(\theta,n)]_{\rm MagMB} \xrightarrow{\ B\to 0 \ }
	H(\theta,n) 2i \sin \frac12 \left({\stackrel{\leftarrow}{\partial}_\theta \stackrel{\rightarrow}{\partial}_n -\stackrel{\leftarrow}{\partial}_n \stackrel{\rightarrow}{\partial}_\theta}\right) \tilde f(\theta,n)\, ,
\end{equation}
which is the expression used in \eqref{eq_mag_moyal}.

\subsection{Semiclassical limit of Landau level wavefunctions}\label{ssec_semiclassics}

We provide a derivation of the semiclassical approximation of Landau level wavefunction used in the main text:%
    \footnote{See \cite{ArovasQHE} for a review on Landau level wavefunctions, and \cite{Nguyen:2016itg} for a discussion of the semiclassical limit. However, capturing subleading density response even at $B \to  0$ will require going beyond the approximations of \cite{Nguyen:2016itg} and involves a triple scaling limit, see below.}
\begin{align}
    \langle n_1,m_1 |e^{-i\vb q\cdot\hat{\vb x}} | n_2,m_2\rangle
	= 
	\langle n_1 |e^{-i\vb q\times \vb k / B} | n_2\rangle 
	\langle m_1 |e^{-i\vb q\cdot \vb R} | m_2 \rangle \, .
\end{align}
where
\begin{align}
    \langle n_1 |e^{-i\vb q\times \vb k / B} | n_2\rangle &= e^{-|q|^2l^2/4} \sqrt{\frac{n_-!}{n_+!}} L_{n_-}^{(n_+ - n_-)}\left( \frac{|q|^2l^2}{2} \right) 
\begin{cases}
    \left( \frac{\bar{q}l}{\sqrt{2}} \right)^{n_+ - n_-} & \qquad \text{if $n_+ = n_2$} \\
    \left( \frac{-q l}{\sqrt{2}} \right)^{n_+ - n_-} & \qquad \text{if $n_+ = n_1$}
\end{cases} 
\end{align}
where $q = q_x + iq_y$,  $l = 1/\sqrt{B}$ is magnetic length, $n_{+} = \max\{n_1, n_2\}, n_{-} = \min\{n_1, n_2\}$, and $L_{\alpha}^{\beta}(z)$ is generalized Laguerre polynomials. We change variable to $n=\frac{n_1+n_2}2$ and $\Delta n = n_2-n_1$. Our goal is to evaluate
\begin{align}
    &\sum_{\substack{\Delta n = -2n\\ \Delta n = 2n \text{  mod  } 2}}^{2n} 
    \mel{n - \frac{\Delta n}{2}}{e^{-i\vb q\times \vb k / B}}{n + \frac{\Delta n}{2}} e^{-in\Delta \theta} \\
    = &\sum_{\substack{\Delta n = 0\\ \Delta n = 2n \text{  mod  } 2}}^{2n} e^{-|q|^2l^2/4}\sqrt{\frac{(n-\Delta n/2)!}{(n+\Delta n/2)!}} \left( \left( \frac{\bar{q}l}{\sqrt{2}} \right)^{\Delta n} e^{-i\Delta n \theta} + \left( \frac{-ql}{\sqrt{2}} \right)^{\Delta n} e^{i\Delta n \theta} \right) L_{(n-\Delta n/2)}^{(\Delta n)}\left( \frac{|ql|^2}{2} \right), 
\end{align}
where $n \in \mathbb{Z}^+/2$. We anticipate that all three argument of the generalized Laguerre polynomial to scale with $n$, which eventually will be set to $\nu_F$. To make the triple scaling manifest, we define $\eta = \Delta n/n$, and $\kappa = |ql|^2/2n$.

Using the following integral representation
\begin{align}
     L_\alpha^{(\beta)}(z) = \frac{1}{2\pi i} \oint_C \frac{e^{-zt/(1-t)}}{(1-t)^{\beta+1} t^{\alpha+1}} dt
\end{align}
we find the $n \rightarrow \infty$ limit of the generalized Laguerre polynomial, in our region of interests, gives
\begin{align}
      L^{(n \eta)}_{n(1-\eta/2)}(n \kappa) &= \frac{e^{-n \Re \phi(t^*)} |f(t^*)|}{\sqrt{2\pi n |\phi''(t^*)|}}  2 \cos \left( n \Im \phi(t^*) - \text{Arg} f(t^*) + \frac{1}{2} \text{Arg} \phi''(t^*) - \frac{\pi}{2} \right),
\end{align}
with 
\begin{align}
    f(t) = \frac{1}{t(1-t)}, \quad \phi(t) = \frac{\kappa t}{1-t} + \eta\log(1-t) + (1-\eta/2) \log(t), \quad t^*_\pm = \frac{2-\kappa \pm \sqrt{\eta^2 - (4-\kappa)\kappa}}{2+\eta}
\end{align}
Using the Stirling approximation $n! \approx \sqrt{2\pi n}e^{n \ln n - n}$, we can obtain the asympototic limit of the square root term. In the end, we find all the exponential factors exactly cancels and are left with with oscillatory terms. Approximating $\sum_{\Delta n} \approx 2 n \int d\eta$, our desired expression becomes
\begin{align}\label{eta_integral}
    4\sqrt{\frac{n}{2\pi}} \int_0^2 d\eta e^{i\frac{\pi}{2}n \eta} \cos\left( (\varphi+\pi/2) n \eta \right) \cos\left(n s_1(\eta) + s_2(\eta) \right) \left( \frac{1}{(4-\kappa)\kappa-\eta^2}\right)^{1/4}
\end{align}
where 
\begin{align}
    s_1(\eta) = \frac{\sqrt{|\Delta|}}{2} + \theta_1(1-\eta/2) - \varphi_1 (1+\eta/2) \qquad  
    s_2(\eta) =\frac{1}{2}(\theta_1 - \varphi_1) - \frac{\pi}{4}
\end{align}
with $\theta_1 = \atan(\sqrt{|\Delta|}/(\eta-\kappa))$ and $\varphi_1 =\atan(\sqrt{|\Delta|}/(\eta+\kappa)) $, and $\Delta = (4-\kappa)\kappa - \eta^2$. The resulting expression is still complicated, but one can perform saddle-point approximation by taking the $n \rightarrow \infty$ limit. In the end, we find Eq. \ref{eta_integral} evaluates to
\begin{align}
     \sqrt{\frac{8-2(\eta^*)^2}{8 -2\kappa -(\eta^*)^2}} \exp{i \left[ N g(\eta^*) + \frac{1}{2}(\theta_1(\eta^*) - \varphi_1(\eta^*)) \right]}
\end{align}
where 
\begin{align}
    (\eta^*)^2 &= \frac{\kappa(4-\kappa) + \kappa^2 \sin^2(2\theta) + \text{sgn}(\cos(2\theta))\sqrt{(\kappa(4-\kappa) + \kappa^2 \sin^2(2\theta))^2 - 16 \kappa^2 \sin^2(2\theta)}}{2}
\end{align}
This is exact in the $n \rightarrow \infty$ limit. In practice, we are interested in small $\kappa = q^2/p_n^2$ expansion, where $p_n = \sqrt{2Bn}$ defines the Fermi surface radius when $n = \nu_F$. Taylor expanding around $\kappa=0$  up to $\mathcal{O}(q^2)$ gives 
\begin{align}
    \sum_{\Delta n} \mel{n - \frac{\Delta n}{2}}{e^{-i\vb q\times \vb k / B}}{n + \frac{\Delta n}{2}} e^{-in\Delta \theta} = \left(1 - \frac{1}{8}\frac{q}{p_n} \cos(2\theta) \right)  e^{-i\frac{2n}{p_n} q \sin(\theta) \left(1 - \frac{1}{24}(2+2\cos(2\theta)) \frac{q^2}{p_n^2} \right) }
\end{align}
where we have used rotation invariance to set $\mathbf{q} = (q, 0)$. One can easily obtain higher order correction in $q$ by keeping additional terms in the Taylor series expansion. For arbitrary $\vb{q}$, the effect is to shift $\theta \rightarrow \theta + \theta_q$ with $\theta_q = \tan^{-1}(q_y/q_x)$. Using $\hat{n}$ to denote direction normal to the Fermi surface, the frame-covariant expression is given by
\begin{align}
    &\sum_{\Delta n} \mel{n - \frac{\Delta n}{2}}{e^{-i\vb q\times \vb k / B}}{n + \frac{\Delta n}{2}} e^{-in\Delta \theta} = \left(1 + \frac{1}{8p_n^2}\left(q^2 - 2(\mathbf{q}\cdot \hat{n})^2  \right)   \right) e^{-i\frac{p_n }{B}(\mathbf{q}\cross \hat{n}) \left[ 1 - \frac{q^2 -  2(\mathbf{q}\cdot \hat{n})^2}{24 p_n^2}\right]}
\end{align}
where we used $2n/p_n = p_n/B$.

\subsection{Geometric corrections to density response}\label{app_q2_rhorho}

In Sec.~\ref{sec_corrections}, we studied corrections to density response coming from a non-parabolic dispersion relation in 2d. These have a similar structure to those arising from a nonlinear (non-Luttinger) dispersion in 1d bosonization. 
Extended Fermi surfaces also have other, inevitable, corrections that encodes the geometry of the Fermi surface. In our approach, these arise from the fact that after abelianizing, the density operator is not linear in the compact bosons, unlike in 1d \eqref{eq_j_nonab}. Let us explain how this comes about:
below Eq.~\eqref{eq_rhoq}, we found that the density operator could be expressed
\begin{equation}\label{eq_rho_interm}
\rho(\vb q)
	= \int \frac{dn d\Delta n d\theta}{2\pi} \langle n-\tfrac{\Delta n}2 |e^{-i\vb q \times \vb k / B} | n+\tfrac{\Delta n}2 \rangle e^{-i\theta \Delta n} \tilde f_{\vb q}(\theta,n)\, , 
\end{equation}
where $\tilde{f_{\vb q}}(\theta,n) = \Tr T_{\vb q} \tilde f(\theta,n)$, and 
\begin{align}
\tilde f(\theta,n)= g \star \tilde f_0 \star g^{-1}
	&= g(x) \left[\tilde f_0(p+ \tfrac{i}{2}\stackrel{\leftarrow}{\partial_x} - \tfrac{i}2\stackrel{\rightarrow}{\partial_x})\right]g^{-1}(x)\\
	&= \Theta(\nu_F - n) + \delta(n-\nu_F) \left[i g \partial_\theta g^{-1}\right] 
	+ \delta'(n-\nu_F) \left[-\frac12 \partial_\theta g \partial_\theta g^{-1}\right] \notag\\ \notag
	&\quad + \delta''(n-\nu_F) \left[-\frac{i}{4!} 
		\left(g \partial_\theta^3 g^{-1} + 3 \partial_\theta^2 g \partial_\theta g^{-1}\right)\right] + \cdots\, .
\end{align}
In App.~\ref{ssec_semiclassics}, we further obtained a semiclassical approximation for the matrix element in \eqref{eq_rho_interm}:
\begin{equation}
\begin{split}
\alpha(\vb q,\theta,\nu_F) 
	&\equiv \int {d\Delta n}  \langle \nu_F-\tfrac{\Delta n}2 |e^{-i\vb q \times \vb k / B} | \nu_F+\tfrac{\Delta n}2\rangle e^{-i\theta \Delta n} \\
    &\simeq \left(1 + \frac{1}{8p_F^2}\left(q^2 - 2(\mathbf{q}\cdot \hat{n})^2  \right)   \right) e^{-i\frac{p_F }{B}(\mathbf{q}\cross \hat{n}) \left[ 1 - \frac{q^2 -  2(\mathbf{q}\cdot \hat{n})^2}{24 p_F^2}\right]}\, .
\end{split}
\end{equation}
Putting everything together, we obtain the following expression for the density operator
\begin{equation}\label{eq_rho_final}
\begin{split}
\rho(\vb q)
	&= \int \frac{d\theta}{2\pi} \, \alpha(\vb q,\theta,\nu_F) \left[g i \partial_\theta g^{-1}\right]_{\vb q}
	+ \partial_{\nu_F}\alpha(\vb q,\theta,\nu_F) \left[\frac12 \partial_\theta g \partial_\theta g^{-1}\right]_{\vb q} \\
	&+ \partial_{\nu_F}^2\alpha(\vb q,\theta,\nu_F) \left[-\frac{i}{4!} \left(g \partial_\theta^3 g^{-1} + 3 \partial_\theta^2 g \partial_\theta g^{-1}\right)\right]_{\vb q} + \cdots \, ,  
\end{split}
\end{equation}
where $[\cdot ]_{\vb q} \equiv \tr \left(T_{\vb q} \cdot\right)$. We choose the $T_{\vb{q}}$ operators to be given by \cite{Dersy:2022kjd}
\begin{align}
    T_{\vb{q}} = T_{(n_1, n_2)} = \alpha^{n_1 n_2/2} Z^{n_1} X^{n_2}, \qquad Z \ket{m} = \alpha^{m} \ket{m}, \quad X \ket{m} = \ket{m+1}
\end{align}
with $\alpha = e^{i2\pi/N_{\Phi}}$, and $Z, X$ are the clock and shift matrices of $U(N)$ and $m, n_1, n_2 = 0, \cdots N_\Phi - 1$. Furthermore, they satisfy $ZX = \alpha XZ$, and one can use them to show that commutation relation of Eq. \eqref{eq_UN_algebra}.

We can now explain the origin of the diagrams in the second line of Fig.~\ref{fig_loop_correction}. Abelianizing, the second term above produces a contribution $(\partial_\theta \phi)^2$ to the density. The third term above produces a $\partial_\theta^3 \phi$, represented by a cross in the second line of Fig.~\ref{fig_loop_correction} (it also produces a $(\partial_\theta\phi)^3$ which does not contribute at one-loop because it is normal ordered).

These geometric corrections can be straightforwardly evaluated. There is a slight subtlety in their scaling, compared to nongeometric corrections. Notice that in our expansion, theta derivatives $\partial_\theta\sim qp_F/B$ are always accompanied by $\partial_\nu$'s. When a $\partial_\nu$ acts on an object like $\epsilon(k^2/2)$, as in Sec.~\ref{sec_corrections}, it scales as $\partial_\nu \sim B/p_F^2$, so that the expansion $\partial_\nu \partial_\theta \sim q/p_F$ corresponds to the desired low wavevector expansion for the $B\to 0$ Fermi liquid. However, in Eq.~\eqref{eq_rho_final}, $\partial_\nu$ is acting on the semiclassical matrix element $\alpha \simeq e^{iq_s p_F / B}$, which has explicit $B$ dependence. In this case, $\partial_\nu \sim q/p_F$ instead of $\partial_\nu \sim B/p_F^2$. It would then seem that $\partial_\nu \partial_\theta \sim q^2 / B$ corrections are singular. One can verify that these vanish---the non-vanishing contributions have an additional $\delta \theta\sim B/(q p_F)$ suppression, such that the final combination $q/p_F$ is finite as $B\to 0$ and corresponds to the expected expansion.

Let us illustrate one of these corrections by considering the first diagram in the second line of Fig.~\ref{fig_loop_correction}. The left-hand side of the diagram comes from the second term in \eqref{eq_rho_final}, whereas the right-hand side involves a single cubic vertex $\partial_\nu^2\epsilon$ from \eqref{eq_delta_S}. Our approach will be to abelianize, with a different choice than the one followed around \eqref{eq_j_ab}: we will now consider the $N_{\Phi}$ generators $T_{m,m'}$ with matrix elements $(T_{m,m'})_{m_1m_2} = \delta_{m,m_1}\delta_{m',m_2}$. These lead to a very simple action in terms of $N_{\Phi}$ decoupled (but self-interacting) bosons:
\begin{equation}
S = 
	-\int \frac{dt d\theta}{4\pi} \sum_m\partial_\theta \phi_m (\partial_t  + \omega_c \partial_\theta)\phi_m
	-\int \frac{dt d\theta}{2\pi}\frac1{3!} \omega_1 \sum_m (\partial_\theta \phi_m)^3 \, ,
\end{equation}
with $\omega_1 \equiv \partial_\nu^2 \epsilon$.
The density operator, to our level of precision, is 
\begin{equation}
\begin{split}
\rho(\vb q)
	&\simeq \int \frac{d\theta}{2\pi} \, e^{i\frac{p_F }{B}q_s}\left(\left[g i \partial_\theta g^{-1}\right]_{\vb q}
	+ i\frac{q_s}{p_F}\left[\frac12 \partial_\theta g \partial_\theta g^{-1}\right]_{\vb q} \right)
	\,.
\end{split}
\end{equation}
where $q_s \equiv \vb{q} \cdot \hat{s} $ and $\hat{s} = \hat{z} \times \hat{n}$ is the unit vector tangential to the Fermi surface, which we typically pick to be $\hat{s} = (-\sin \theta, \cos \theta)$. 
Abelianizing gives
\begin{align}
\left[g i \partial_\theta g^{-1}\right]_{\vb q}
	&= \sum_{m,m'} j_{m,m'}(\theta) \tr \left(T_{\vb q} T_{m, m'}\right)
	= \alpha^{\frac{n_{q_x} n_{q_y}}{2}} \sum_{m} j_{m, m-n_{q_y}}(\theta) e^{im \psi_{q_x}}\\
\left[\partial_\theta g \partial_\theta g^{-1}\right]_{\vb q}
	&= \sum_{m_1, m_1', m_2, m_2'} :j_{m_1, m_1'}(\theta) j_{m_2, m_2'}(\theta): \tr \left(T_{\vb q} T_{m_1, m_1'}T_{m_2, m_2'}\right)
	\\
    &= \alpha^{\frac{3n_{q_x} n_{q_y}}{2}}\sum_{m_1, m_2} :j_{m_1, m_2+n_2}(\theta) j_{m_2, m_1}(\theta): e^{i m_2 \psi_{q_x}}
\end{align}
with $g^{-1}i \partial_\theta g^{-1} = \sum_{m, m'} j_{m,m'}(\theta) T_{m,m'}$ with $j_{m, m'}(\theta) \sim :e^{i(\phi_m(\theta) - \phi_{m'}(\theta))}:$ and $j_{m,m}(\theta) = i\partial_\theta \phi_m(\theta)$ is understood to be the diagonal current. We also used $\psi_{q_x} = \frac{2\pi }{N_{\Phi}} \frac{q_xL}{2\pi} = 2\pi \frac{n_{q_x}}{N_{\Phi}}$ ($n_{q_x} \equiv \frac{q_x L}{2\pi} = 0,\ldots , N_{\Phi}-1$), and that $\langle m |T_{\vb q}| m'\rangle = \alpha^{n_{q_x} n_{q_y}/2} \alpha^{m n_{q_x}} \delta_{m, m'+n_{q_y}} $. The phases $e^{im \psi_q}$ will cancel in density correlator below, but it will play a role in Landau parameter corrections. For two-point function of density, we align external $\vb{q}$ along the $q_x$ direction, which sets $n_{q_y} = 0$. We also drop the total $U(1)$ current in the subleading term since it doesn't contribute to density response at finite $q,\omega$. The density simplifies to
\begin{equation}
\rho(q)
	\simeq \int \frac{d\theta}{2\pi} \, e^{i\frac{k_F }{B}q_s}\sum_m e^{im \psi_q} \left(\partial_\theta \phi_m + i \frac{q_s}{2 p_F} (\partial_\theta \phi_m)^2 \right)
	\,.
\end{equation}
Define the Fourier transform of $\phi_m(t,\theta) = \sum_\ell \int \frac{d\omega}{2\pi}e^{i\ell\theta - i\omega t} \phi_{m}(\omega, \ell)$. We find 
\begin{align}
    \langle \phi_m \phi_{m'} \rangle(\omega, \ell) = \frac{\delta_{m,m'}}{\ell(\omega - \omega_c \ell)}
\end{align}
We are often interested in correlation function of $\rho_m = \partial_\theta \phi_m$, given by
\begin{align}
    \langle \rho_m \rho_{m'} \rangle(\omega, \ell) = \delta_{m,m'} \frac{i \ell}{\omega - \omega_c \ell}, \qquad \langle \rho_m(t,\theta) \rho_{m'}(0,0) \rangle = \delta_{m,m'}\frac{1}{(\omega_c t - \theta)^2}
\end{align}
The $\omega_1$ correction to density two-point function gives
\begin{equation}
\begin{split}
\delta \langle \rho(q) \rho(-q) \rangle(\omega) = \int \frac{d\theta d\theta'}{(2\pi)^2}dt & e^{i\omega t}e^{-i\frac{k_F}{B}q(\sin \theta - \sin \theta')} \sum_{m,m'} e^{i(m-m')\psi_q} \frac{iq}{2p_F}\\
&\quad \times \left(\sin(\theta) \langle \rho_m^2 (iS_{\rm int}) \rho_{m'} \rangle - \sin(\theta') \langle \rho_m (iS_{\rm int}) \rho_{m'}^2 \rangle \right)
\end{split}
\end{equation}
where
\begin{align}
    \langle \rho_m^2 (iS_{\rm int}) \rho_{m'} \rangle &= \frac{-i \omega_1}{3!} \int_X \sum_{m_1} \langle \rho_m^2(A) \rho_{m_1}^3(X) \rho_{m'}(B) \rangle \notag\\
    &=  \delta_{m,m'}\frac{-i\omega_1}{3!} \int_X 6 G_{AX}^2 G_{XB}  \\
    &= -i\delta_{m,m'}\omega_1 [G^2 * G](t-t', \theta-\theta')  \notag
\end{align}
where $A=(t, \theta), B = (t', \theta'), X = (t_1, \theta_1)$, and $G = \frac{1}{(\omega_c t - \theta)^2}$ with $*$ denote convolution. Therefore, we find
\begin{align}
     \delta \langle \rho(q) \rho(-q) \rangle(\omega) &= N_{\Phi}\omega_1 \frac{q}{2p_F} \int \frac{d\theta d\theta'}{(2\pi)^2} dt e^{i\omega t -i\frac{k_F}{B}q(\sin \theta - \sin \theta')} \left(\sin(\theta) - \sin(\theta') \right) [G^2 * G](t-t', \theta-\theta')\notag  \\
     &= N_{\Phi}\omega_1 \frac{q}{2p_F} \int \frac{d\bar\theta d\delta \theta}{(2\pi)^2} dt e^{i\omega t - i\frac{k_F}{B}q\cos(\bar\theta)\delta \theta} \cos(\bar\theta)\delta \theta[G^2 * G](t-t', \delta \theta)
\end{align}
Using convolution theorem, we find
\begin{align}
    \mathcal{F}[G*G^2] = \widetilde{G}\widetilde{G^2} = \frac{1}{3!}\frac{\ell^4}{(\omega - \omega_c \ell)^2}
\end{align}
with $\widetilde{G}, \widetilde{G^2}$ being their respective Fourier transform. Let $\ell = \frac{k_F}{B}q\cos(\bar{\theta})$. It follows that
\begin{align}
    \delta \langle \rho(q) \rho(-q) \rangle(\omega) &= N_\Phi \omega_1 \frac{q}{2p_F} \int \frac{d\bar\theta}{2\pi} \cos(\bar \theta) (i\partial_\ell) \left(\frac{1}{3!}\frac{\ell^4}{(\omega - \omega_c \ell)^2} \right) \notag\\ \notag
    &= i N_\Phi \omega_1 \frac{q}{2p_F} \int \frac{d\bar\theta}{2\pi} \cos(\bar \theta) \frac{q^3 v_F^3 \cos^3(\bar\theta)}{3\omega_c^3 \left(\omega - v_F q \cos(\bar{\theta})\right)^3} \left( 2 \omega - v_F q \cos(\bar{\theta}) \right) \\
    &= i N_\Phi \omega_1 \frac{q^2}{p_F} \frac{v_F}{\omega_c^3} \int \frac{d\bar\theta}{2\pi} \frac{\cos^4(\bar\theta)}{6\left(s-\cos(\bar\theta)\right)^3}\left(2s - \cos(\bar \theta)\right) \\ \notag
    &= i N_\Phi \omega_1 \frac{q^2}{p_F} \frac{v_F}{\omega_c^3} \left( \frac{4s^3 - s^5}{12(s^2-1)^{5/2}} + \frac{1}{12} \right)
\end{align}
Dividing by the volume factor and rearranging the leading coefficient gives
\begin{align}
    \delta \langle \rho \rho \rangle(\omega, q) = i \frac{p_F}{2\pi v_F} \frac{q^2}{p_F^2} \frac{\omega_1 p_F^3}{B^2 v_F} \left( \frac{4s^3 - s^5}{12(s^2-1)^{5/2}} + \frac{1}{12}\right)
\end{align}
where we used $N_\Phi = \frac{BV}{2\pi}, \omega_c = \frac{B}{p_F/v_F}$. The scaling function in parentheses agrees with the fermion result shown in Eq.~\eqref{eq_fermion_rr_final}, matching $4g_1 + 2g_2$ with the identification $\omega_1 = \epsilon''/B^2$. This demonstrate the validity of our bosonization framework in computing loop corrections to 2D Fermi liquid theory.

\subsection{Landau Parameter Correction}\label{app_LP_corr}
We show that the leading Landau parameter action indeed reproduces the the correction to density response from Landau parameters. Starting with action
\begin{align}
    S_{\rm int} = \frac{\omega_c}{2 N_{\Phi}}\int dt \frac{d\theta d\theta'}{(2\pi)^2} \sum_{\vb{q}} e^{-i\vb{q}\times (\vb{k_F}(\theta) - \vb{k_F}(\theta'))/B} j_{\vb{q}}(t, \theta) j_{-\vb{q}}(t, \theta') F_{\vb{k}_F(\theta), \vb{k}_F(\theta')}
\end{align}
with $j_{\vb{q}} = \frac{1}{2\pi} \tr(gi\partial_\theta g^{-1}T_{\vb{q}})$ satisfying
\begin{align}
    \langle j_{\vb{q}}(t, \theta) j_{\vb{q}'}(t', \theta') \rangle = \frac{N_{\Phi}\delta_{\vb{q},-\vb{q}'}}{(\omega_c (t-t') - (\theta - \theta'))^2} 
\end{align}
where we have decompactified the $\theta$ coordinate. We choose the external momentum to align in the x-direction $\vb{q} = (q, 0)$. The correction to connected density response due to Landau parameters is given by
\begin{align}
    \delta \langle \rho (q) \rho(-q) \rangle(\omega) &= -i\frac{\omega_c}{2N_{\Phi}} \int \frac{d\theta d\theta'}{(2\pi)^2} dt e^{i\omega t-i\frac{k_F}{B}(\sin \theta - \sin \theta')} \int dt_1 \frac{d\theta_1 d\theta_2}{(2\pi)^2} \sum_{\vb{q}_1} e^{i \vb{q} \cdot (\hat{s_1}-\hat{s_2})/B} F(\theta_1 - \theta_2) \\
    & \langle j_{\vb{q}}(t, \theta) j_{\vb{q}_1}(t_1, \theta_1) j_{\vb{q}_1}(t_1, \theta_2) j_{\vb{-q}}(t=0, \theta') \rangle \\
    &= -i \omega_c N_{\Phi} \int \frac{dt d\theta d\theta_1}{(2\pi)^2}  \frac{e^{i\omega(t-t_1)} e^{-i\frac{k_F}{B}q(\sin \theta - \sin \theta_1)}}{(\omega_c (t-t_1) - (\theta - \theta_1))^2}  \\
    &\int \frac{dt_1 d\theta_2 d\theta'}{(2\pi)^2}  \frac{e^{i\omega t_1} e^{-i\frac{k_F}{B}q(\sin \theta_2 - \sin \theta')}}{(\omega_c t_1 - (\theta_2 - \theta'))^2} F(\theta_1 - \theta_2) \\
    &= i \frac{N_{\Phi}}{\omega_c} \int \frac{d\bar{\theta} d\bar{\theta}'}{(2\pi)^2} \frac{v_F q \cos \bar\theta}{\omega-v_F q \cos \bar{\theta}} F(\bar{\theta} - \bar{\theta}') \frac{v_F q \cos \bar\theta'}{\omega-v_F q \cos \bar{\theta}'} 
\end{align}
where we defined $\bar{\theta} = (\theta + \theta_1)/2, \bar{\theta}' = (\theta_2 + \theta' )/2$. Dividing by the volume factor gives
\begin{align}
    \delta \langle \rho \rho \rangle(\omega, q) = i \frac{p_F}{2\pi v_F}\int \frac{d\bar{\theta} d\bar{\theta}'}{(2\pi)^2} \frac{v_F q \cos \bar\theta}{\omega-v_F q \cos \bar{\theta}} F(\bar{\theta} - \bar{\theta}') \frac{v_F q \cos \bar\theta'}{\omega-v_F q \cos \bar{\theta}'} 
\end{align}
which is indeed the leading correction to density due to Landau parameters.

\section{Free Fermi gas observables}\label{app_FFG}

\subsection*{Specific heat}

For a free Fermi gas, it is simplest to obtain the thermal partition function from the Fock space spectrum:
\begin{equation}
\begin{split}
\log Z 
	= \log \Tr e^{-\beta (H-\mu Q)}
	&=\log \sum_{\{n_k\}} e^{-\beta \sum_k \left(\epsilon_k - \mu n_k\right)}\\
	&= \log \prod_k \sum_{n_k} e^{-\beta \left(\epsilon_k - \mu n_k\right)}\\
	&= \sum_k \log [1+e^{-\beta(\epsilon_k - \mu)}]]\\
	&= V \int \frac{d^dk}{(2\pi)^d} \log [1+e^{-\beta (\epsilon_k - \mu)}]\, .
\end{split}
\end{equation}
It is equal to $\beta V P$. Its derivative with respect to $\beta$ is UV-finite:
\begin{equation}\label{eq_heat}
\mathcal E - \mu n 
	= P - sT
	\equiv- \partial_\beta \frac{1}{V} \log Z
	= \int_k \frac{\epsilon_k - \mu}{1+e^{\beta(\epsilon_k-\mu)}}
	\equiv \int_k (\epsilon_k - \mu) f_{\rm FD}(\epsilon_k - \mu)\, .
\end{equation}
It becomes simpler to compute after taking a second derivative with respect to $\beta$, which gives the specific heat:
\begin{equation}\label{eq_cV_fermion_calc}
\begin{split}
c_V = \frac{d (\mathcal E - \mu n )}{dT}
	= \beta^2 \partial_\beta^2 \frac{1}{V} \log Z
	&= \int_k \left[\frac{\frac12\beta (\epsilon_k - \mu)}{\cosh\frac12\beta(\epsilon_k-\mu)}\right]^2\\
	&\simeq \int_{-\infty}^\infty d\xi \underbrace{\frac{S_{d-1}}{(2\pi)^d}\frac{k^{d-1}(\xi)}{\epsilon'_k(\xi)}}_{\nu(\xi)} 
	\left[\frac{\frac12\beta\xi}{\cosh\frac12\beta\xi}\right]^2\, ,
\end{split}
\end{equation}
where in the last step we changed variables to $k\to \xi = \epsilon_k - \mu$, and dropped exponentially small terms $\sim e^{-\beta\mu}$ by taking the lower limit of integration $-\mu \to -\infty$. At low temperatures, the integrand is sharply peaked around $\xi = 0$, so that we have 
\begin{equation}
c_V
	\simeq \underbrace{\frac{S_{d-1}}{(2\pi)^d} \frac{k_F^{d-1}}{\epsilon'_k(0)}}_{\equiv \nu(0)} T \int ds \left[\frac{\frac12s}{\cosh\frac12s}\right]^2
	= \frac{\pi^2}{3} \nu(0) T\, .
\end{equation}
If the single-particle density of states $\nu(\xi)$ is constant, i.e. $\epsilon_k \propto k^d$, then the result above is exact (up to terms exponentially suppressed at low temperature). When $\nu(\xi)$ is not constant, we can expand it around $\xi = 0$ to obtain
\begin{align}\label{eq_cV_general_d}
\frac{1}{\frac{\pi^2}{3} \nu(0) T}c_V
	&= 1+\frac{7\pi^2}{10}  \frac{\nu''(0)}{\nu(0)} T^2 
		+ \frac{31\pi^4}{168} \frac{\nu''''(0)}{\nu(0)} T^4  + \cdots\\
	&= 1 + \frac{7}{10} \left(\frac{\pi T}{\epsilon' k_F}\right)^2 \left[(d-1)(d-2) - 3 (d-1)\frac{k_F \epsilon''}{\epsilon'}  + 3 \left(\frac{k_F \epsilon''}{\epsilon'} \right)^2 - \frac{k_F^2 \epsilon'''}{\epsilon'} \right] + \cdots  \, . \notag
\end{align}
%

\subsection*{Dynamical spectral function}

In 2d, the density two-point function of free Fermi gas with parabolic dispersion $\epsilon(k) = \frac{k^2}{2m}$ is $\langle \rho \rho \rangle = i \Pi$ with \cite{PhysRevB.58.15449}
\begin{align}
    \Pi (\omega, q) &= \frac{1}{2\pi}\left(-1 + \frac{p_F}{|q|}\left( \sqrt{\left(-is+\frac{|q|}{2p_F}\right)^2 -1} + \sqrt{\left(is+\frac{|q|}{2p_F}\right)^2 -1} \right) \right) \, , 
\end{align}
with $s = \omega/(v_F q)$. The small-$q$ expansion is
\begin{align}
\frac{v_F}{p_F}\Pi (\omega, q)
	&= \frac{1}{2\pi}\left(-1+ \frac{s}{ \sqrt{s^2-1}}\right)  + \frac{1}{16\pi }\frac{s}{(s^2-1)^{5/2}} \left(\frac{q}{p_F}\right)^2 + \mathcal{O}(q^4)
\end{align}
To establish further checks of our approach, we generalize this result to fermions with arbitrary dispersion relation. In general, the polarization tensor is given by
\begin{align}
\langle \rho \rho \rangle (\omega, q) = i\int \frac{d^2 k}{(2\pi)^2} \frac{\Theta(k_F - k) - \Theta(k_F - |k+q|)}{\omega - (\epsilon_{k+q} - \epsilon_k)}
\end{align}
Let $\vb q = (q, 0)$ and write $k_x = k \cos(\theta), k_y = k \sin(\theta)$. We note that 
\begin{equation}
|k+q| = \sqrt{k^2+q^2 + 2k q \cos(\theta)} = k + q \cos(\theta) + \frac{q^2 \sin^2(\theta)}{2k} - \frac{q^3 \cos(\theta) \sin^2(\theta)}{2k^2} 
\end{equation}
and
\begin{equation*}
\begin{split}
\Theta(k_F - k) - \Theta(&k_F - |k+q|) = \delta(k_F-k) \left( \cos(\theta) q + \frac{\sin^2(\theta) q^2}{2k} - \frac{q^3 \cos(\theta) \sin^2(\theta)}{2k^2} \right) \\
    &+ \frac{1}{2}\delta'(k_F-k) \left(\cos^2(\theta) q^2 + \frac{q^3 \cos(\theta) \sin^2(\theta)}{k} \right) + \frac{1}{3!} \delta'''(k_F-k) q^3 \cos^3(\theta) 
\end{split}
\end{equation*}
Expanding both the numerator and denominator in powers of $q$ gives
\begin{equation}\label{eq_rr_fermion_gen}
-i \frac{2\pi v_F}{p_F} \langle \rho \rho \rangle (\omega, q)
	= g(s) + \frac{q^2}{p_F^2} \left( g_0(s)  + \frac{\epsilon''p_F}{v_F}  g_1(s)  + \left(\frac{\epsilon'' p_F}{v_F}\right)^2 g_2(s)  + \frac{\epsilon'''p_F^2}{v_F}  g_3(s)\right) + \cdots
\end{equation}
with
\begin{equation}
\begin{split}
	g(s)	&= -1 + \frac{s}{\sqrt{s^2-1}}\\
    g_0(s) &=  \frac{2 s^2 \left(\sqrt{s^2-1}-s\right)+s}{8 \sqrt{s^2-1}} \\
    g_1(s) &= \frac{1}{24} \left(1+ 6 s^2+\frac{8 s^3-6 s^5}{\left(s^2-1\right)^{3/2}} \right)  \\
    g_2(s) &= \frac{1}{24} \left(-1-12 s^2+\frac{\left(12 s^4-29 s^2+20\right) s^3}{\left(s^2-1\right)^{5/2}}\right) \\
    g_3(s) &= g_1(s)\,.
\end{split}
\end{equation}
In Eq.~\eqref{eq_rr_fermion_gen}, derivatives of the dispersion are taken with respect to $k$, e.g.~$\epsilon''=\partial_k^2\epsilon(k)|_{k=k_F}$. To simplify the comparison with the bosonization results in Sec.~\ref{sec_corrections}, where filling $\nu = \frac{k^2}{2B}$ is the natural variable, it will be convenient to change variables to $d(\nu B) = d(k^2 /2) = k dk$. This amounts to the replacement $\epsilon'\to k \epsilon' $, $\epsilon''\to \epsilon' + k^2 \epsilon''$, $\epsilon'''\to 3k \epsilon'' + k^3 \epsilon'''$. In terms of these derivatives, the result becomes:
\begin{equation}\label{eq_fermion_rr_final}
-i \frac{2\pi v_F}{p_F}  \langle \rho \rho \rangle (\omega, q) = g + \frac{q^2}{p_F^2} \left( 
	g_0 + g_1 + g_2  
	+ \frac{\epsilon'' p_F^3}{v_F} \left(4g_1 + 2 g_2\right)  
	+ \left(\frac{ \epsilon'' p_F^3}{v_F}\right)^2  g_2  
	+ \frac{\epsilon''' p_F^5}{v_F}  g_1 \right) \, .
\end{equation}

\bibliographystyle{ourbst}
\bibliography{qboso}{}

\end{document}